\documentclass[8.5pt,twoside]{article}

\usepackage{amsmath,mhchem}
\usepackage{color}
\usepackage{xcolor}
\usepackage[super,sort&compress,comma]{natbib} 
\usepackage{times,mathptm}
\usepackage{sectsty}
\usepackage{balance} 
\usepackage{multirow}
\usepackage{array}
\usepackage[text={11.3cm,20.4cm},centering]{geometry} 

\usepackage{graphicx} 
\usepackage{lastpage}
\usepackage[format=plain,justification=raggedright,singlelinecheck=false,font=small,labelfont=bf,labelsep=space]{caption} 
\usepackage{fancyhdr}
\begin{document}

\thispagestyle{plain}
\fancypagestyle{plain}{
\fancyhead[L]{\includegraphics[height=8pt]{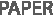}}
 \fancyhead[C]{\hspace{-1cm}\includegraphics[height=15pt]{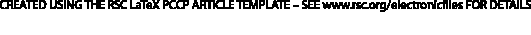}}
 \fancyhead[R]{\includegraphics[height=10pt]{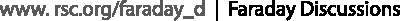}\vspace{-0.2cm}}
\renewcommand{\headrulewidth}{1pt}}
\renewcommand{\thefootnote}{\fnsymbol{footnote}}
\renewcommand\footnoterule{\vspace*{1pt}%
\hrule width 11.3cm height 0.4pt \vspace*{5pt}} 
\setcounter{secnumdepth}{5}

\makeatletter 
\renewcommand{\fnum@figure}{\textbf{Fig.~\thefigure~~}}
\def\subsubsection{\@startsection{subsubsection}{3}{10pt}{-1.25ex plus -1ex minus -.1ex}{0ex plus 0ex}{\normalsize\bf}} 
\def\paragraph{\@startsection{paragraph}{4}{10pt}{-1.25ex plus -1ex minus -.1ex}{0ex plus 0ex}{\normalsize\textit}} 
\renewcommand\@biblabel[1]{#1}            
\renewcommand\@makefntext[1]%
{\noindent\makebox[0pt][r]{\@thefnmark\,}#1}
\makeatother 
\sectionfont{\large}
\subsectionfont{\normalsize} 

\fancyfoot{}
\fancyfoot[LO,RE]{\vspace{-7pt}\includegraphics[height=8pt]{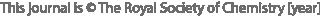}}
\fancyfoot[CO]{\vspace{-7pt}\hspace{5.9cm}\includegraphics[height=7pt]{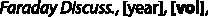}}
\fancyfoot[CE]{\vspace{-6.6pt}\hspace{-7.2cm}\includegraphics[height=7pt]{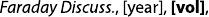}}
\fancyfoot[RO]{\scriptsize{\sffamily{1--\pageref{LastPage} ~\textbar  \hspace{2pt}\thepage}}}
\fancyfoot[LE]{\scriptsize{\sffamily{\thepage~\textbar\hspace{3.3cm} 1--\pageref{LastPage}}}}
\fancyhead{}
\renewcommand{\headrulewidth}{1pt} 
\renewcommand{\footrulewidth}{1pt}
\setlength{\arrayrulewidth}{1pt}
\setlength{\columnsep}{6.5mm}
\setlength\bibsep{1pt}

\noindent\LARGE{\textbf{Quantum chemistry, classical heuristics, and quantum advantage}}
\vspace{0.6cm}

\noindent\large{\textbf{Garnet Kin-Lic Chan$^{\ast}$}}\vspace{0.5cm}

\noindent\textit{\small{\textbf{Received Xth XXXXXXXXXX 20XX, Accepted Xth XXXXXXXXX 20XX\newline
First published on the web Xth XXXXXXXXXX 200X}}}

\noindent \textbf{\small{DOI: 10.1039/c000000x}}
\vspace{0.6cm}

\noindent \normalsize{We describe the problems of quantum chemistry, the intuition behind classical heuristic methods used to solve them, a conjectured form of the classical complexity of quantum chemistry problems, and the subsequent opportunities for quantum advantage. This article is written for both quantum chemists and quantum information theorists. In particular, we attempt to summarize the domain of quantum chemistry problems as well as the chemical intuition that is applied to solve them within concrete statements (such as a classical heuristic cost conjecture and a classification of different avenues for quantum advantage) in the hope that this may stimulate future analysis.}
\vspace{0.5cm}

\section{Introduction}

This essay is about the quantum simulation of chemical matter. Although this is the introductory article to a Faraday Discussion in a chemistry journal, it is actually written with two audiences in mind: quantum chemists and quantum information theorists. 
This is because although quantum chemistry and quantum information theory have increasingly crossed paths
in recent years, practitioners in one field often have limited understanding of the other field's perspective. 
One purpose of this essay is to describe the intuition that quantum chemists have about the quantum many-body problem in chemical matter. This intuition guides modern day research into improved methods and their applications. The other is to give a vantage point on quantum chemistry that hopefully emphasizes some of the concerns of quantum information theorists, which we believe will be useful for the future development of quantum chemistry. Quantum information theory is a mathematical field with provable results, while quantum chemistry is  primarily an empirical one. As the author is a quantum chemist,  this article is written in the informal style of quantum chemistry. In some cases, it supplies the author's (non-rigorous) personal opinions. Intuitions and opinions are clearly not theorems, but we present them in the hope that they can be valuable guideposts when the path ahead is unclear. 

This essay has four subsequent parts. First, we give a brief tour of  quantum mechanics (QM) in chemical matter and review some applications of modern day interest. Second, we summarize a range of classical heuristics in quantum chemistry. Third, we give an intuitive perspective on how to think about complexity in quantum chemistry. Finally, we briefly discuss quantum algorithms for quantum chemistry and where to search for quantum advantage in the simulation of chemical matter.

\subsection{An apology about discussions of complexity}

 This article makes statements about complexity in the non-rigorous quantum chemistry style. We use terms such as polynomial (easy) and exponential (hard) complexity, and mention some complexity classes {such as \textsc{NP} (harder than polynomial to solve (e.g. exponentially hard) but polynomially easy to verify) and \textsc{QMA} (the analog of \textsc{NP} on a quantum computer)~\cite{arora2009computational}. But we go no deeper. We hope that by understanding the intuitions that quantum chemists have about their problems,  computer scientists may begin to formalize the conjectures about complexity that this article raises.

\section{Quantum mechanics in chemical matter simulation}

\subsection{What is quantum chemistry?}

We first consider the term ``quantum chemistry''. Unfortunately this term has
grown to have different meanings in chemistry and in quantum information theory, which is a source of confusion. In the quantum information and quantum computing communities, quantum chemistry usually means ``an application of quantum mechanics to chemistry''. However, in the chemical sciences, quantum chemistry has traditionally had a more restricted meaning: it is the numerical and conceptual theory of chemical bonding and reactivity. Mathematically this is respresented by the  electronic structure problem. In theoretical chemistry,  quantum chemistry is a sub-field that is considered to be distinct from quantum dynamics and quantum statistical mechanics.

A second source of confusion is that in quantum information discussions, chemistry usually means molecular chemistry, perhaps related to the popular view of chemistry as something to do with brightly colored solutions bubbling away in test-tubes. Yet in the chemical sciences, chemistry has long grown beyond only the molecular setting to encompass materials and biochemistry. In this broadened perspective, the defining characteristic of chemistry is the understanding of matter at the level of the chemical identity of the atoms, under the relatively accessible conditions of a chemical laboratory.


As quantum chemistry and quantum information come together, it is perhaps most interesting, and most useful, to use a broader definition of the term quantum chemistry, as the application of quantum mechanics to molecules, materials, and biomolecules under relatively accessible experimental conditions, which we collectively refer to here as chemical matter. In this essay, we will start with this broad definition of quantum chemistry. However, for reasons of scope and the expertise of the author, we will necessarily need to specialize and we focus only on the electronic structure aspects of  quantum chemistry later on.

\subsection{Quantum mechanics in chemistry}

So how does quantum mechanics enter into chemistry? Chemical matter consists of electrons (with mass $\sim 10^{-30}$~kg) and nuclei (with masses in the range $\sim 10^{-27}-10^{-25}$~kg) and these particles are typically confined on the atomic scale (about $10^{-10}$~m). Because of their very different masses, at a given energy we can think of the motion of electrons as being much faster than that of the nuclei, so that they quickly reach some kind of stationary state. Thus the starting point for much of quantum mechanics in chemistry is the Born-Oppenheimer approximation. This is an adiabatic (i.e. separation of timescales) approximation, i.e. for a given set of nuclear positions $\mathbf{R}$, we can meaningfully discuss the electronic eigenstates.
These are the eigenstates of the electronic Hamiltonian \begin{align}
  {H}^\text{el} = {T}^\text{el}+V^\text{nuc-el} + V^\text{el-el} + V^\text{NN}
  \end{align}
containing the kinetic energy of the electrons ($T^\text{el}$), the nuclear-electron potential ($V^\text{nuc-el}$), the electron-electron repulsion ($V^\text{el-el}$), and the nuclear-nuclear repulsion $V^\text{NN}$, and we write the time-independent electronic Schr\"odinger equation as
\begin{align}
  H^\text{el}( \mathbf{R})  \Psi_i^\text{el}( \mathbf{r};  \mathbf{R} ) = E_i^\text{el}  \Psi_i^\text{el}( \mathbf{r};  \mathbf{R} )
  \end{align}
where $\mathbf{r}$ are the electron coordinates, and we indicate a parametric dependence of the electronic wavefunctions and energies on the nuclear positions $\mathbf{R}$. $E_i^\text{el}( \mathbf{R})$ is known as the potential energy surface (PES) of the $i$th electronic state and is the foundation for the most basic concepts of molecular structure: the minima define the stable molecular structures and the saddle points define the transition states which govern reaction rates.
Because of the central nature of the PES, we can reasonably argue that electronic structure  is the defining mathematical problem of quantum mechanics in chemistry and is where the majority of applications of quantum mechanics in chemistry lie.

Electronic structure is, however, not the only way in which quantum mechanics enters into chemistry. For example, the time-independent  electronic Schr\"odinger equation has its time-dependent counterpart
\begin{align}
  i \partial_t \Psi^\text{el}(\mathbf{r};\mathbf{R}) = H^\text{el} \Psi^\text{el}(\mathbf{r};\mathbf{R}),
\end{align}
(here and elsewhere we work in atomic units) and the associated problem of electron dynamics enters into problems of electron transport, energy transfer, and strongly (e.g. laser) driven processes in chemistry.

In addition, the nuclei in the Born-Oppenheimer approximation may be thought of as particles moving in the effective classical potential that is the potential energy surface $E^\text{el}_i( \mathbf{R})$. The stationary quantum solutions then satisfy the time-independent nuclear Schr\"odinger equation
\begin{align}
  (T^\text{nuc} + E^\text{el}_i) \Phi^\text{nuc}_{i, \alpha}(\mathbf{R}) = E^\text{tot}_{i, \alpha} \Phi_{i, \alpha}^\text{nuc}(\mathbf{R})
  \end{align}
Solving for the energies $E^\text{tot}_{i, \alpha}$ and $\Phi_{i, \alpha}$ is the basis of understanding various kinds of spectroscopy which excite the nuclear motion. The time-dependent nuclear Schr\"odinger equation follows as
\begin{align}
  i \partial_t \Phi^\text{nuc}(\mathbf{R}) = \left(T^\text{nuc} + E^\text{el}_i\right) \Phi^\text{nuc} (\mathbf{R})
  \end{align}
and describes nuclear quantum effects in reaction dynamics. And finally, there are quantum mechanical efforts outside of the Born-Oppenheimer approximation. These are needed to understand the non-adiabatic processes of relaxation of electrons, spins, and nuclei, as well as phenomena where there is an electronic energy scale comparable to the nuclear vibrational energy scale, such as in phonon-driven superconductivity. 


\subsection{The magnitude of quantum effects in chemistry: classical limits}

The above categorizes the subareas of quantum mechanical effects in chemistry, but does
not speak to how important they are. As Dirac famously pointed out, quantum mechanics is the underlying theory of chemistry~\cite{dirac1929quantum}. But in practice, many chemical phenomena can be understood without, or only with a little, quantum mechanics. An estimate of the size of quantum effects in different settings is from the thermal de Broglie wavelength $\lambda = \sqrt{\frac{2\pi}{mkT}}$; this is the lengthscale on which the delocalized nature of quantum particles starts to affect their statistical properties. At room temperature (298~K), for electrons, $\lambda_e = 10^{-9}$~m and for the nuclei it ranges from $\lambda_\text{nuc} \sim 10^{-10} - 10^{-12}$~m. At the atomic scale of $10^{-10}$~m, we conclude that quantum effects are always important in describing electrons, but are usually unimportant for nuclei, which are essentially classical objects, with the exception of the lightest nuclei.

\subsection{The magnitude of quantum effects in chemistry: locality}

Another important empirical limitation to the emergence of quantum effects in chemistry is the ``principle'' of locality. It is a famous feature of quantum mechanics that states can carry non-local correlations (entanglement), in principle, over long distances. But chemical thinking succeeds precisely by assuming that such effects do not exist: we can reason about chemical entities because distant parts of a molecule do not radically change the identity of a given atom. 

As we discuss somewhat more technically in Sec.~\ref{sec:locality}, the existence of locality in chemistry can reasonably be argued to be partly due to the laws of Nature, and partly due to the lack of perfect quantum control over chemical matter, which limits the types of systems that are studied in chemical experiments. Whatever the origin, the locality principle limits the complexity of quantum effects: coherence and entanglement are limited to finite distances and finite times. {The task of computational quantum simulations is to address quantum effects up to these finite scales.}

\subsection{Standard computational applications of QM in chemistry}

Because of the above, in practice, the use of QM to understand chemistry is more restricted than Dirac's original statement suggests. Some examples of ``standard'' computational applications of QM in chemistry include (i) computing the potential energy surface by solving the electronic Schr\"odinger equation with some approximate method, (ii) computing rates associated with dynamical processes of electrons, photoexcitations, and energy relaxation, often using low-order time-dependent perturbation theory, (iii) describing corrections to classical nuclear motion, commonly in a semi-classical picture.

The motivation for new methods in quantum chemistry naturally arises due to limitations in our capabilities in the above tasks. As all methods are approximate, the limitations are usually either of the kind that the approximation is not accurate enough, or the approximation is not cheap enough!
 These are clearly two sides of the same coin, but as many heuristics have a setting that one can adjust to balance cost versus accuracy, the first problem refers to the case where dialing up the accuracy leads to us exceeding our computational budget, while the second refers to the case where dialing down the accuracy still does not lead to a fast enough calculation (or else produces nonsense). 


Discussing the methods and the limitations in more detail for all the above tasks exceeds the scope of this essay and the capabilities of this author. In the remainder, we focus primarily on the task of solving the electronic Schr\"odinger equation, the original domain of quantum chemistry.


\subsection{The nature of open problems in chemistry}

Many computational chemistry problems remain open. This simply means that they cannot be solved at a level that advances chemical understanding. 
At the same time, there is an important difference when applying  computation to chemistry, such as to answer ``what is the mechanism of nitrogenase'', as compared to a problem such as ``find the prime factors of 43,112,609''. 
One difference is that the chemical problem is much less well posed (e.g. what should we even compute?) A second difference is that 
there is a lot of existing partial information, from  experimental observation, by analogy with other related chemical systems, and from a range of inexact classical simulations. Chemical simulations are thus always about adding the next chapter to an ongoing story, rather than opening and closing the book. When assessing the value of improved simulations and new methods in quantum chemistry, the fact that open problems are not really completely open constrains the role of new simulations in the field.

\subsection{Some examples of open problems}

\label{sec:openproblems}

We now consider some examples of open computational chemistry problems that have been used to motivate the development of new quantum chemistry methods, namely (i) the computation of ligand affinities in drug discovery, (ii) understanding biological nitrogen fixation, and (iii) ambient pressure high temperature superconductivity.

\subsubsection{Ligand binding affinities in drug discovery }
 Drug discovery is a long and complicated task, costing upwards of 1 billion USD to develop  a single new small molecule drug, thus the possibility to use relatively inexpensive computational chemistry to improve this process is an alluring possibility. Although most stages of the drug pipeline do not involve quantum mechanics, the basic thesis in computational drug design is that a necessary (though obviously far from sufficient) requirement is that the drug molecule attaches to its target (usually a protein). Thus, a seemingly obvious task for a quantum chemical simulation would be to compute the binding affinity of a drug molecule at a given protein target site.

Assuming that the protein is relatively rigid and the target site and general orientation (``pose'') of the drug are already known, computing the binding affinity, however, is still quite non-trivial. Simple approximate mean-field quantum chemistry methods - such as density functional theory (DFT, see Sec.~\ref{sec:hfdft}) - in most cases yield sufficient accuracy in the electronic energy, and a density functional calculation of the energy can indeed today be carried out on small proteins and drugs~\cite{ryde2016ligand}. However, the challenge arises from the fact that the binding affinity is not a zero-temperature energy, but a free energy, that includes the entropic effects from the solvent and the protein. {In actual binding events, the entropy and energy effects are often similar in magnitude with opposite sign, with the binding free energy being the small residual~\cite{gilli1994enthalpy}.} The motion of the atoms can be assumed to be entirely classical, described by Newton's equations, 
but in standard methodologies (such as  free-energy perturbation theory which computes the relative affinities of ligands relative to a base ligand) one needs to simulate, even under optimistic assumptions, $\sim 100$~ns of dynamics, or roughly $10^{8}$ energies and forces. For this to complete in a realistic turnaround time (say 1 day of compute) each electronic energy and gradient must be computed in $\sim 1$ millisecond.

Thus computing ligand affinities is a problem where standard QM simulations, as applied directly to the protein plus ligand complex, are simply not fast enough. Instead, practical solutions for free energy sampling involve empirical models trained to the QM data: these either abandon the QM description entirely for a parametrized ``force-field'' form of the potential energy surface (perhaps, today, enhanced with a general function approximator such as a neural network); or else simulate only a small part of the problem near the binding site quantum mechanically (and the rest with force fields); these methods can be combined with statistical mechanical ``enhanced sampling'' techniques to reduce the number of atomic configurations to be computed for the free energy. 
The goal of developing new quantum methods in this area (such as faster DFT algorithms or even cheaper mean-field methods) is to partially bridge the gap to the point that training techniques have sufficient data; then they can be hoped to generalize accurately to the timescales of interest. 

\subsubsection{Biological nitrogen fixation } Another problem often brought up in the context of improving quantum chemistry methods is that of understanding biological nitrogen fixation. The nitrogen cycle is one of the major biogeochemical cycles and the fixing (i.e. reduction) of atmospheric nitrogen is essential for the biological incorporation of nitrogen. Nitrogen is fixed today both
by the Haber-Bosch process, an industrial process nominally given by the equation
\begin{align}
\ce{N_2} + \ce{3H_2} \to  \ce{2 NH_3}
\end{align}
which balances the kinetic and thermodynamic driving forces of the process at high pressures and high temperatures and uses an iron-based catalyst, and by bacteria via the nitrogenase enzyme, which works under ambient conditions, nominally satisfying the stoichiometry
\begin{align}
\ce{N2} + \ce{8 H+} + \ce{8 e^-} + \ce{16 MgATP} \to \ce{2 NH3} + \ce{H2} + \ce{16 MgADP} + \ce{16 P_i}
  \end{align}

One oft-quoted statistic about the Haber-Bosch process is that it consumes an enormous amount of energy, e.g. 2 percent of the world's energy supply. Efficient Haber-Bosch plants today use about 26~GJ per ton of ammonia~\cite{ROUWENHORST202141}. It is tempting to think that biological nitrogen fixation uses much less energy, but  this is not simple to establish. For example, the above equation suggests 16 ATP molecules are consumed for every \ce{N2} molecule, translating under physiological ATP hydrolysis conditions (which can vary, leading to slightly different numbers~\cite{milo2015cell}) to a nominal 24~GJ per ton of ammonia, a comparable energy cost. But  
this is not an end-to-end cost, e.g. it does not include the costs of
synthesizing and maintaining the nitrogenase machinery (a significant component of the cellular proteome in nitrogen fixing cells)
as well as general cellular function and homeostasis. In addition, there are other challenges besides energy consumption in industrializing a biological process, for example, the nitrogenase enzyme is extremely slow.
The chemical motivation to study nitrogenase is thus less to produce an energy-efficient replacement of the Haber-Bosch process but rather because it is an interesting system in its own right, and perhaps it may motivate how to understand and design other catalysts that can activate and break the nitrogen-nitrogen triple-triple bond under ambient conditions.
{In this context, some basic chemical questions to answer include (i) where do the various species in the reaction bind to the enzyme, and (ii) what is the step-by-step catalytic mechanism?}


The focal point of the earliest quantum chemistry calculations has mainly been to determine the electronic structure of the Fe-S clusters that are the active sites of nitrogenase, which would seem to be the starting point to understand their catalytic activity. This problem is non-trivial because the Fe atoms contain 3d orbitals and the associated electrons are ``strongly correlated'' (discussed in detail in Sec.~\ref{sec:strongcorr}), posing a practical challenge for many common quantum chemistry approximations. At the current time, there is a reasonable understanding of the general qualitative features of the electronic structure of the iron-sulfur clusters from a combination of quantum chemistry calculations and experimental spectroscopy~\cite{sharma2014low,li2019electronic,bjornsson2015discovery}. However, the computational challenge of treating the strongly correlated electronic structure at a quantitative level means that the precise nature of the ground-state of the important FeMo cofactor (where nitrogen reduction takes place) is not fully resolved.


Computing the electronic ground state, although a useful reference point, is still of limited value in deciphering the mechanism. More directly useful, while still computationally accessible, quantities are the zero-temperature relative energetics e.g. to determine proton binding sites and protonation states of the Fe-S clusters and their environments. This requires simulating not just the isolated Fe-S clusters but also the surrounding residues, a simulation scale of a few hundred atoms~\cite{jiang2023protonation}. The only practical quantum chemistry method for this task currently is density functional theory. Although it is well known
that DFT can yield erroneous predictions in transition metal systems, one can still obtain useful information by looking for consistent results between different DFT approximations and constraining the results by experimental data. There is great need for improvement in density functionals (or other methods) that can be applied at the scale of these problems which will impact our understanding here. Looking beyond zero-temperature energies to free energies, there are monumental challenges to overcome similar to those discussed for ligand binding free energies, but now with the additional complications of the non-trivial electronic structure.

\subsubsection{Ambient pressure high-temperature superconductivity }

As our last example, we consider the problem of ambient pressure high-temperature superconductivity, as exemplified by the cuprate materials. Cuprates involves strongly correlated electron physics driven by the narrow Cu $3d$ electron bands. Here there are a plethora of questions, ranging from the mechanism underlying superconductivity to the nature of different exotic phases observed in these materials. In the chemical setting, a particularly relevant question is how to design new material structures and compositions to exhibit higher superconducting transition temperatures.

While many mysteries of high temperature superconductors remain unexplained, it is important to recognize that after almost four decades of work, there is substantial partial understanding from both theory and experiment. Limiting oneself only to ground-states and static properties, in the context of simple models such as the 1-band (Fermi)-Hubbard model, numerical calculations nowadays give a fairly comprehensive view of the potential ground-states, although in some parts of the phase diagram it is not entirely clear what the true ground-state order is due to close competition between different orders~\cite{zheng2017stripe,arovas2022hubbard,xu2024coexistence}. {We also note that none of the proposed ground-states are of some qualitatively ``classically intractable variety'' (i.e. the orders still have a succinct classical description, see Sec.~\ref{sec:classicalstateprep})}; but the uncertainty arises because the precise refinement of their energies requires too large a cost. {For the zero-temperature states, in the author's view achieving a 10$\times$ improvement in accuracy over the current state-of-the-art would likely resolve most outstanding questions about this energetic competition.} 

In the context of material specific predictions, by neglecting some of the physics it is also becoming possible to perform concrete calculations of the low-temperature properties of different material structures. Both parametrized Hamiltonians~\cite{weber2010strength,schmid2023superconductivity} as well as fully multi-orbital ab initio approaches~\cite{cui2022systematic,cui2023ab} have been deployed for this task, and predictions can be made of modest quality. The techniques involved are extensions of the simulation methods used with simpler models, with a polynomial increase in computational complexity associated with the increased material realism. On the other hand, in practice one still needs to make choices about what physics to include and what to neglect. Much work remains to improve all aspects of these simulation methods in terms of speed, accuracy, and generality (e.g. to simultaneously include correlated electron, phonon, and finite temperature effects).  
We are still quite far from being able to make precise predictions about T$_c$ in calculations that cover all microscopic mechanisms of potential interest.

\subsection{Summary}

We can summarize our brief introduction to quantum effects in chemistry and our survey of problems as follows. The first is 
that quantum mechanics is not necessary to understand all of chemistry, but in the cases where it is needed, there are indeed many open problems which challenge current classical approximate quantum chemistry methods, both in terms of the speed required and in terms of accuracy. From a complexity perspective, this often appears to involve either large constant prefactors (e.g. in the problem of sampling free energies) or (as we argue in more detail later, see Sec.~\ref{sec:classicalrefinement}) polynomial factors e.g. when seeking improved accuracy in strongly correlated problems. At the same time, the classical approximate methods we have, combined in some cases with experimental knowledge, are already remarkably successful across a range of difficult problems. New algorithms and new simulation tools must compete successfully with these existing heuristics to have an impact in the chemical setting.

\section{Classical heuristic quantum chemistry}

\label{sec:heuristics}

\subsection{Energy scales in the electronic Schr\"odinger equation}

\label{sec:energies}

We now turn to an overview of classical electronic structure methods in quantum chemistry. These are necessarily approximate methods, and they either implicitly or explicitly assume some features of the solution. Thus we also refer to these approximations as classical heuristics. We assume that the computational task is to solve for the ground- and/or low-lying electronic states, where low-lying could mean e.g. thermally accessible, or accessible by applying some perturbation, such as a laser.

The (non-relativistic) electronic Hamiltonian is written in atomic units as
\begin{align}
  H = \sum_i \left(-\frac{1}{2}\nabla_i^2 +  v_\text{ext}(\mathbf{r}_i) \right)+ \sum_{i>j} \frac{1}{r_{ij}} + E_{NN} \label{sec:Hamiltonian}
\end{align}
where we have dropped the ``el'' superscript, $v_\text{ext}$ is the field external to the electrons (which in the absence of an applied field, is the nuclear-electron potential $v^{\text{nuc-el}}$) and
$E_{NN}$ is, for the purposes of determining the electronic wavefunction at a given nuclear geometry, a constant, which we drop when convenient in the discussions below. We will be interested in eigenstates with $N$ electrons, where $N$ makes the system charge neutral (or close to neutral).
Of course the 1-electron part of the Hamiltonian leads to trivial eigenstates, and it is the 2-electron part of the Hamiltonian (the Coulomb repulsion) that leads to correlated electronic structure, the central source of complexity. This general structure of the Hamiltonian and the energy scales of the 1-electron and 2-electron terms in different systems sets up the qualitative aspects of the electronic structure of the ground- and low-lying states.

As basic intuition, we first consider the energy scales of electrons in atoms. Atomic orbitals are defined as eigenstates of an (effective) 1-particle Hamiltonian (see mean-field methods below, Sec.~\ref{sec:hfdft}), and we typically divide the orbitals into core, valence, Rydberg, and continuum states. The core states lie at an energy of $O(Z^2)$ (where $Z$ is the nuclear charge); valence states at energies of $O(1)$; Rydberg states at close to $0$, and the continuum states at positive energies.

Chemistry mainly involves the valence orbitals. As atoms come together in a bond (in a single-particle mean-field description) the valence orbitals interact with each other and their energies split, and the range of splittings takes a range of values, say {$O(0.01-1)$ (atomic units, i.e. Hartrees)} in chemically relevant systems; we might denote this energy scale as $t$. The Coulomb matrix elements of the valence orbitals (on the same atom) are typically in the range of {$O(0.1-1)$} (atomic units); we can denote this $U$. Thus the effect of the Coulomb interactions on the mean-field picture of electronic structure can be understood in terms of the ratio of these energies $U/t$; for large $U/t$ the Coulomb interaction is non-perturbative, or ``strong'', while for small $U/t$ it is a perturbation, or ``weak''. We emphasize that in a material, it is the full range of splittings, not an individual energy level splitting such as the bandgap, which determines if the Coulomb interaction should be thought of as strong. This is because the number of electrons immediately around the bandgap (or Fermi surface) is vanishingly small compared to the total number of electrons and does not determine the overall electronic structure. The correct ratio is thus that of the Coulomb interaction to the bandwidth, which is $O(U/t)$.

The implication of the weak and strong limits of the Coulomb interaction is that they lend themselves to different kinds of approximations. Note that strong does not necessarily mean qualitatively complicated, because non-perturbative effects may be captured by a modified starting point, such as a different mean-field theory. We return to this in Sec.~\ref{sec:hfdft} and in Sec.~\ref{sec:classicalstateprep}.

\subsection{Basis sets}

Moving to a computational discussion, classical numerical algorithms (other than some quantum Monte Carlo algorithms, briefly discussed in Sec.~\ref{sec:survey}) do not directly target the continuum Hamiltonian in Eq.~\ref{sec:Hamiltonian}. Instead one first solves the electronic Schr\"odinger equation in a discrete Hilbert space associated with a single particle basis $\{ \chi_p(\mathbf{r}) \}$. The $N$-electron wavefunction lives in the antisymmetric product space of the single-particle basis. 

Fortunately, in the low-lying electronic states in chemical matter, the electrons are largely confined. This greatly simplifies basis construction because (i) we know where the electrons are spatially localized and (ii) we know the feature size of the electrons.

The most popular basis functions have historically been chosen to simplify  the required Hamiltonian matrix elements in the basis. They include atomic basis sets (commonly Gaussian basis sets) and plane wave basis sets. In Gaussian bases, a linear combination of Gaussian functions are centered on each nucleus of the problem, i.e.
\begin{align}
  \chi(\mathbf{r}) = (x-A_x)^i (y-A_y)^j(z-A_z)^k e^{-\alpha (\mathbf{r}-\mathbf{A})^2}
\end{align}
where $\mathbf{A}$ is the position of the nucleus and $i, j, k$ are integers. In addition to knowing the nuclear positions $\mathbf{A}$, (i) and (ii) above mean that  $\max(i, j, k)$ can be restricted to be small, and the range of $\alpha$ limited. These atomic-centred bases offer precise control over the location and shape of the basis functions but are non-orthogonal, which leads to some numerical complications.

Plane wave bases are constructed by placing the system of interest in a finite box of volume $V$ (say cubic, for simplicity) with periodic boundary conditions. This suggests a natural basis function of the form
\begin{align}
  \chi_\mathbf{G}(\mathbf{r}) = \frac{1}{\sqrt{V}}e^{-i \mathbf{G}\cdot \mathbf{r}}
  \end{align}
with the wavevectors $\mathbf{G}$ chosen to satisfy the boundary conditions. Here, (i) means that $G_\text{max}$ is limited. Plane waves uniformly resolve space and are orthogonal, which leads to well-conditioned numerical algorithms with systematic convergence, although they are not as compact as atomic bases.

Regardless of the choice of single-particle basis, it is the sharpest features of the function to be represented that dominate the asymptotic rate of convergence. If the function is not infinitely differentiable, {then in general the error of a  basis expansion will converge algebraically with the number of basis functions $M$.}  In the eigenstates of the Hamiltonian, non-differentiable features arise when the Hamiltonian is singular.  
For mean-field theories, such as Hartree-Fock theory and density functional theory (Sec.~\ref{sec:hfdft}) where the goal is to determine the single particle orbitals $\phi(\mathbf{r})$, the sharp features are the orbital cusps at the nuclei, which are not infinitely differentiable. Plane wave expansions then converge algebraically, but because in a Gaussian basis the functions can be centered on the nucleus, and  the exponents can be chosen in a non-uniform manner, superalgebraic convergence can be achieved.

In the many-body wavefunction however, there is another cusp in the inter-electronic coordinate due to the electron-electron Coulomb singularity at coincidence. The presence of this cusp leads to a convergence with respect to basis size $M$ that is  {$\sim O(1/M)$} regardless of the choice of single-particle basis function above (for a longer discussion that gathers some of the known results, see Appendix E of Ref.~\cite{babbush2018low}). This slow rate of convergence is known as the basis set problem, and remedies to this problem, in the form of introducing analytic functions with an explicit dependence on the inter-electron coordinate, are known as explicit correlation methods~\cite{kong2012explicitly}.

Because electrons are identical particles, it is convenient to rewrite the Hamiltonian in terms of second quantization. Once we have chosen a basis (and assuming the basis has been orthogonalized e.g. in the case of Gaussians), the Hamiltonian becomes
\begin{align}
  H = \sum_{ij} t_{ij} a^\dag_i a_j + \frac{1}{2}\sum_{ijkl} v_{ijkl} a^\dag_i a^\dag_j a_l a_k \label{eq:secondquant}
\end{align}
where $t_{ij}$, $v_{ijkl}$ are referred to as the 1- and 2-electron integrals, and $a^\dag$, $a$, are the electron creation and annihilation operators.

\subsection{Mean-field methods}

\label{sec:hfdft}
With a numerical representation of $H$, we now turn to how the low-lying eigenstates are computed. The starting point for most classical quantum chemistry methods (and most chemical intuition) uses a simple product form for the eigenstates,
\begin{align}
  |\Phi\rangle = \prod_{m=1}^N c_{m}^\dag |0\rangle \label{eq:mfansatz}
\end{align}
where the creation operators create the molecular orbitals  $\{ \phi_m (\mathbf{r})\}$, rather than the computational basis functions above, i.e.
\begin{align}
    c_m^\dag = \sum_{i} C_{mi} a^\dag_i \label{eq:mocoeff}
\end{align}
The simple product wavefunction $|\Phi\rangle$ is usually called a Slater determinant (named after the first quantized form of its amplitudes). The ansatz Eq.~\ref{eq:mfansatz} and methods to determine the molecular orbitals are referred to in quantum chemistry as simply ``mean-field theory'', even though this is of course only one possible kind of mean-field theory.

Conceptually, the simplest mean-field theory is Hartree-Fock theory. Here, the molecular orbitals are determined by variationally minimizing the energy $E=\langle \Phi|H |\Phi\rangle / \langle \Phi|\Phi\rangle$, which leads to a non-linear eigenvalue problem, the Hartree-Fock equations. In an orthogonal basis this takes the form
\begin{align}
  \mathbf{F}(\mathbf{C}) \mathbf{C} = \mathbf{C} \epsilon
  \label{eq:hfeq}
\end{align}
where $\mathbf{F}$ is the one-electron Fock operator, $\mathbf{C}$ contains the molecular orbital coefficients appearing in Eq.~\ref{eq:mocoeff}, and $\epsilon$ are the molecular orbital energies (which we already referred to in our intuitive discussion of energy scales in Sec.~\ref{sec:energies}).

Another common mean-field theory, closely related to Hartree-Fock theory, is Kohn-Sham density functional theory. Because all information in a Slater determinant (other than its absolute phase) is contained in the single particle density matrix $\gamma_{ij} = \langle \Phi |a^\dag_i a_j |\Phi\rangle$, the Hartree-Fock approximation to the energy is a functional of the single-particle density matrix; $E \equiv E[\gamma]$. Density functional theory takes this one step further and writes the {ground-state} energy as a functional of the single-particle real-space density, $\rho(\mathbf{r}) = \sum_{ij} \gamma_{ij} \phi_i (\mathbf{r})\phi_j(\mathbf{r})$; $E \equiv E[\rho]$. 
Intuitively, this functional dependence is achieved because it is valid only for the ground-state, and because the set of electronic Hamiltonians can be labelled by the single particle potentials {$v_\text{ext}(\mathbf{r})$, a label of the same dimensionality as $\rho(\mathbf{r})$.} 

In the Kohn-Sham version of density functional theory, which is the most widely applied form (and what people mean by DFT without further qualification) this energy is broken into several components
\begin{align}
E[\rho] = T_s[\rho] + V_{ne}[\rho] + J[\rho] + E_{xc}[\rho] \label{eq:ksdft}
\end{align}
where the first 3 terms (the Kohn-Sham kinetic energy, nuclear-electron attraction, and classical Coulomb repulsion) are known in a computationally efficient form, but $E_{xc}[\rho]$ (the exchange-correlation functional) is not. Eq.~\ref{eq:ksdft} is an identity, so one can take the perspective that DFT is exact, one only needs to know $E_{xc}[\rho]$, but of course that offers no computational simplification, and in practice $E_{xc}[\rho]$ is some numerical approximation that gives the resulting DFT simulation the flavour of mean-field theory. The minimization of $E[\rho]$ is achieved by solving the Kohn-Sham equations for the orbitals which constitute the density. The Kohn-Sham equations in a basis take the same form as the Hartree-Fock equations in Eq.~\ref{eq:hfeq}.


The cost of Hartree-Fock and DFT calculations can be summarized as the cost to evaluate the energy in a basis (which is related to the cost of computing the Hamiltonian matrix elements) together with the cost to minimize (to a local minimum or stationary point) the Hartree-Fock or DFT energy. The most common way to find a minimizing solution is to solve the Hartree-Fock or Kohn-Sham non-linear eigenvalue problem. Depending on the basis representation, building the Fock matrix can be done in $O(N^3)-O(N^4)$ time, while solving the Fock equations can be performed as a set of self-consistent diagonalizations, each of which takes $O(N^3)$ time. The number of iterations is unknown and this procedure only guarantees a stationary point of the energy; formally finding the global minimum, being a non-convex optimization, is computationally \textsc{NP} hard. 

The formal \textsc{NP} hardness of mean-field optimization, however, provides an important example of the difference in  intuition offered by worst case complexity and the practical experience of quantum chemists. In practice, we have a lot of information about the kinds of solutions we are looking for. For example, the principle of locality (see Sec.~\ref{sec:locality}) suggests that it is useful to assemble a guess of the Hartree-Fock solution from solutions of atomic Hartree-Fock calculations, and starting from such guesses often provides quick convergence to a local minimum. In large systems, locality also reduces the cost in other ways, due to the sparsity of the Hamiltonian in a local basis. And of course, one can argue that large physical systems cannot reach their global minimum anyways, if such a minimum is hard to find. 

\subsection{The role of mean-field theories in quantum chemistry}

The mean-field description of the low-lying eigenstates is conceptually simple and attractive. In quantitative calculations also, the modeling of any problem almost always starts with a mean-field calculation. This is in part testament to the power of empirical density functional parametrizations which can achieve the remarkable accuracy of a few kcal/mol on standardized (albeit limited in diversity) small molecule thermochemistry datasets~\cite{mardirossian2017thirty}. This level of accuracy is often enough to inform experimentalists, for example, to distinguish between reaction pathways, or predict the correct products.
But even in the cases where Hartree-Fock and density functional methods fail, they still form the starting point for more complex calculations. This is mainly for two reasons. 

The first is that often in real chemical applications, even if one is interested in only the zero-temperature energy differences, multiple electronic structure calculations must be performed to optimize the nuclear positions and explore the potential energy surface (see e.g. Sec.~\ref{sec:openproblems}). Mean-field methods therefore enter as an initial, if imperfect, guide to this landscape. 

The second reason is arguably deeper, and is related to the existence of  different types of mean-field solutions.
The non-convexity of the Hartree-Fock (or approximate Kohn-Sham) energy optimization means that 
multiple low-energy local minima can exist. As every Slater determinant is a type of classical state, in the sense that it has no entanglement in the molecular orbital basis, this means that there are many potential ``classical'' approximations to the quantum ground-state. 
In general, the local minimizing Slater determinants need not transform as an irreducible representation of the symmetry group of $H$. When they do not, we say the mean-field solution is a broken symmetry solution. But in the thermodynamic limit, the true ground-state may indeed also break symmetry, a phenomenon known as spontaneous symmetry breaking. (In fact, symmetry breaking characterizes many of the common electronic phases of matter, such as ferromagnets, antiferromagnets, superconductors and others). 


Consequently, although a given mean-field calculation may not provide a reasonable description of the system of interest, the set of low-energy  mean-field states may still contain an appropriate qualitative classical description among them.  Selecting the appropriate mean-field reference recovers a simple and intuitive picture of the eigenstate that may be the starting point for more advanced treatments.
We return to this below.  

\subsection{Qualitative electron correlation: weak fluctuations}

Moving beyond the mean-field description of quantum chemistry, we now consider how to include electron correlation. The mean-field Slater determinant is an eigenstate of the Fock operator, thus correlation consists of the fluctuations generated by $H-F$. For quantum chemistry calculations that start from a mean-field starting point, we can 
use similar intuition to that expressed in Sec.~\ref{sec:energies}, replacing $t$ by $F$, and $U$ by $W=H-F$, to distinguish between weak and strong correlations.


Weak correlations are sometimes referred to as ``dynamic'' correlations. On the atomic scale, they arise as fluctuations of occupancy between orbitals that are widely separated in energy (such as the valence and the continuum). On longer scales, they also include fluctuations involving orbitals which are spatially widely separated (and thus weakly coupled).

Although individually such fluctuations are small they can add up to a large effect, for example, by integrating over the continuum of orbitals to excite to, or over a large domain in space. In practice, dynamic correlations are thus always quantitatively important. 
{A traditional rule of thumb is that dynamic correlation introduces about 1 eV of correlation energy per pair of electrons in an occupied orbital~\cite{jensen2007}.} 
Excitations to high energy continuum states can modify the wavefunction on  small length scales and thus help to capture features near the singularity of the electron-electron cusp in the wavefunction. Because of the need to describe excitations to the continuum, dynamic correlation requires large basis sets to capture faithfully.   

\subsection{Qualitative electron correlation: strong fluctuations}

\label{sec:strongcorr}

Strong fluctuations from the mean-field reference lead to strong mixing between different valence electron configurations  of the mean-field orbitals. For this reason, strong correlation is also known as multireference correlation. Its presence signals a breakdown of the original mean-field description. 
The loss of an essentially classical starting point might seem to leave one at the mercy of the full complexity of the quantum many-body problem. However, in practice, one encounters one of several simplifications in actual chemical matter. (We note, of course, that even if these simplifications indeed cover all scenarios, this does not render all problems tractable in practice). 

The first is that the degeneracy of valence electron configurations (on the scale of the fluctuations) may only involve a few orbitals. This is encountered in chemical processes such as bond breaking, where the stretching of a given bond causes the valence orbitals associated only with that bond to become strongly correlated, or in a molecular photoexcitation, where only specific orbitals are involved in the excitation process, or in a finite magnetically coupled molecule, with a finite number of coupled spins. Since only a small number of orbitals are involved in any given bond or the excitation or in the spin couplings, this limits the possible degeneracy.

The second simplification occurs when there is a large number of nearly degenerate valence orbitals in a material, as occurs for example, when there are $d$ or $f$ orbital bands. One might worry that one needs an asymptotically exponentially large number of configurations to mix, but this is a scenario where it is useful to consider the entire set of low-energy mean-field solutions, including the broken symmetry solutions. 
If there is only a small (i.e. non-exponentially growing) number of low-energy mean-field broken-symmetry solutions, then for many problems one can assume that the true ground-state in fact is described by one of them, spontaneously breaking symmetry. For example, imagine that for a lattice translationally invariant (crystal) Hamiltonian, one finds a set of translational symmetry breaking mean-field states at low energy. As the system size increases, if these states differ by more and more fluctuations because they correspond to spatial electronic patterns which are globally displaced with respect to each other, then they will not couple via a finite power of the Hamiltonian in the thermodynamic limit, and one can choose any one of them as the starting point to describe the eigenstate. (Should one wish to restore the translational invariance, one can apply the translation operator to the symmetry broken state, generating a linear, i.e. not exponentially large, number of additional configurations). There is an intrinsic length-scale introduced by the symmetry breaking mechanism: once the system size is sufficiently large that the coupling between different broken-symmetry solutions is smaller than the energy accuracy we care about (or our control over the energy of the state in experiment), typically $O(kT)$, then we might as well choose to describe the system with a symmetry broken mean-field state. Then once the appropriate mean-field state is chosen, one can view the remaining correlation as weak correlation.

A third simplification applies even where there is an exponential degeneracy (or near-degeneracy) in the (broken symmetry) mean-field configurations.
While this type of exponential degeneracy  is not much encountered in systems that are  currently of most interest in chemistry, we discuss it here because it does arise in real chemical matter, particularly in the context of magnetism, where it is referred to as frustration. The main simplification that appears here is that even when there is frustration, correlations in chemical matter remain quite local. This makes it possible to represent the low-energy eigenstates with techniques that forgo any mean-field starting point but build in locality, for example the tensor networks discussed in Sec.~\ref{sec:survey}.

\subsection{Qualitative electron correlation: locality}

\label{sec:locality}

Locality is a central simplifying concept in chemical matter. In terms of correlations, we might define a system with (spatial) locality as one where quantum correlation functions $\langle O(\mathbf{r}_1) O(\mathbf{r}_2)\rangle - \langle O(\mathbf{r}_1)\rangle \langle O(\mathbf{r}_2)\rangle$ decay rapidly with $r_{12}$. {In cases where the ground-state has a gap to excitations, we expect these correlations to decay like $e^{-O(r_{12})}$, while in critical systems, we can observe algebraic decay, $O(r_{12}^{-\alpha})$ for positive $\alpha$, e.g in the density matrix elements of a metal~\cite{ismail1999locality} or the spin-correlation of an antiferromagnet~\cite{sandvik1997finite} (associated respectively with gapless modes).} 

The vast majority of problems studied in chemical matter are gapped, even if only because they have a finite size. But it is important to note that even in critical systems, the correlations still decay at long distances, and thus are not as non-local as they could potentially be.

Another way to characterize quantum correlations is via the (bipartite) entanglement entropy $S$ of the state. This counts the (logarithm) of the number of quantum degrees of freedom that are entangled across a spatial dividing surface (the ``boundary''). In terms of entanglement entropy, locality is associated with an area law of entanglement entropy, $S \sim O(L)$, where $L$ is the area of the dividing surface (i.e. only degrees of freedom near a surface can be entangled with the degrees of freedom on the other side). Critical systems show a small correction to the area law, $S \sim O(L \log L)$~\cite{eisert2010colloquium}.  

The main intuition for locality  is simple, namely it stems from the (electronic) Hamiltonian operator, which in a local basis, is quite local. For example, taking the second quantized expression in Eq.~\ref{eq:secondquant}, the one-electron matrix elements in an exponentially localized basis decay exponentially with the separation between basis functions $i$ and $j$. The two-electron matrix elements that derive from the Coulomb interaction decay, somewhat slowly like $1/r_{12}$, but in electrically neutral matter, the presence of the nuclear charges on average eliminates the longest range effects. 
We are then interested in the low-energy eigenstates of the Hamiltonian with this local interaction. Because the ground-state can be obtained as a function of the Hamiltonian (think about the projector onto the ground-state $e^{-\beta H}$ as $\beta \to \infty$), or so the intuition goes, then the low-energy eigenstates inherit this locality in their correlations.

But the above argument is not rigorously right and proving locality of ground-states for different classes of Hamiltonians without additional strong assumptions is notoriously difficult. To see where the difficulty might arise, consider the ground-state as $\lim_{\beta \to \infty} e^{-\beta H}|\Phi\rangle$ (where $|\Phi\rangle$ is some arbitrary state). $\beta$ is here imaginary time, which is naively analogous to an inverse temperature, so we might think that the ``temperature'' need only be significantly below the gap to the first excited state. But for a system of size $L$, even if we assume that the Hamiltonian has a constant gap between the ground and first excited state (i.e. independent of $L$) then one needs  $\beta \sim O(L)$ to avoid the orthogonality catastrophe, as the overlap of $|\Phi\rangle$ with the true ground-state $|\Psi_0\rangle$ is  $e^{-O(L)}$. Then, because the Hamiltonian is being applied $O(L)$ times (e.g. in a polynomial approximation to $e^{-\beta H}$),  these repeated applications could in principle generate longer and longer correlations. 

Regardless of this formal difficulty, chemical intuition is that for most problems we can view the ground-state of a chemical system as indistinguishable from that of a system at a low finite temperature (and, of course, all experiments are performed at a finite temperature). More precisely, this means we really want to model the canonical density operator $\Gamma = e^{-\beta H}$ at some large but system-size independent $\beta$ (i.e. a fixed low temperature), and the empirical finding is that this will not yield very different properties from the ground-state. The use of a fixed, rather than system-size dependent, $\beta$ is consistent with locality, but the statement that $e^{-\beta H}$ and the ground-state yield similar predictions for observables is only true if there is not some very large (exponential in the thermodynamic limit) number of excited states lying just above the ground-state (which would otherwise dominate due to entropy). (For formal connections between the density of states and area laws, see e.g. Ref.~\cite{brandao2015entanglement}). This is violated in some frustrated quantum problems, such as quantum spin glasses, but these are not the typical systems encountered in chemical matter.

Ultimately, we can make a case that locality is prevalent in chemical matter for the combination of reasons that the electronic Hamiltonian is naturally local, and because of a lack of precise control: we do not actually simulate the ground-state, nor do we commonly control chemical experiments and matter sufficiently well, to model Hamiltonians where the low-temperature thermal state and ground-state yield different predictions. Precise quantum control of atoms and molecules is improving in the context of quantum computing. These intuitions about chemical matter may thus need to change in the future if and when the nature of chemical experiments also changes.

Finally, we observe that the locality of quantum correlations does not itself mean that solving the electronic structure problem to find the ground state can immediately be restricted to the cutoff length associated with the correlations alone. This is because even if there are no quantum correlations at all, there can still be non-trivial classical correlations over long distances. 

As a simple example, we can consider a classical Ising model with antiferromagnetic interactions on a geometrically frustrated lattice (e.g. one with triangles). Then the ground-state is a classical state where the connected correlation functions all vanish, i.e. it is some arrangement of up and down spins, that is simply a product state. But we cannot find this product state just by satisfying antiferromagnetic constraints around each triangle (since they cannot be simultaneously satisfied, i.e. it is frustrated). Instead, finding the ground-state requires looking over longer length scales to minimize the frustration. In the worst case, encountered in classical spin glasses, one needs to consider the exponential space of configurations. 

Regardless, in many kinds of chemical matter, finding the ground-state locally in some finite subregion (e.g. measuring observables by simulating a portion of a molecule) does yield results compatible with simulating the ground-state problem as a whole. We can refer to this deeper manifestation of locality as strong locality. In the case of strong locality, the need to model quantum effects is entirely limited to the subregion itself.

\subsection{Evaluating classical heuristics}

Encapsulating the above intuitions within classical algorithms for quantum chemistry is essentially the job of classical heuristics for electron correlation. (Note that there are other simplifying structures in quantum systems which we did not describe above, for example, those associated with sign-free Hamiltonians). There is a very long list of classical quantum chemistry correlation methods, and we only cover some of the most important ones in the sections below. However, in understanding such methods (and when devising new ones), it is useful to analyze them with respect to a few different axes:
\begin{itemize}
    \item  do they capture weak (dynamic) correlation, strong (multireference) correlation, or both, 
    \item are they systematically improvable, 
    \item is the energy extensive (i.e. does one obtain meaningful predictions for a large number of atoms or in the thermodynamic limit of materials), 
    \item what is the cost scaling as a function of system size and as a function of accuracy, 
    \item what is the absolute cost (i.e. what are the computational prefactors?), 
    \item is the method applicable to lattice models only or can it be used with realistic Hamiltonians?
\end{itemize}

\subsection{A brief tour of approximate classical quantum chemistry methods}
\label{sec:survey}

\subsubsection{Configuration interaction }

\begin{table}\centering
\begin{tabular}{c|c}
\hline\hline
      Weak and strong & Both \\
       Systematically improvable & Yes \\
       Extensive energy & Only with exponential cost \\
       Cost scaling & Poly. to exp. depending on truncation  \\
       Prefactors & Very low \\
       Lattice models and ab initio & Both \\
     \hline\hline
\end{tabular}
\caption{Characteristics of configuration interaction methods.
\label{tab:ci}}
\end{table}

This is one of the oldest electron correlation methods in quantum chemistry~\cite{helgaker2013molecular}. It represents the ground-state as a linear combination of different Slater determinants,
\begin{align}
  |\Psi\rangle = \sum_I c_I |\Phi_I\rangle
\end{align}
where the coefficients $c_I$ are determined by diagonalizing the Hamiltonian projected into the space $\{ \Phi_I \}$. If all determinants are included, the method is called full configuration in quantum chemistry and exact diagonalization in physics communities.

The characteristics of configuration interaction methods are listed in Table~\ref{tab:ci}. 
Today, it is commonly used in molecular applications to describe strong correlation in valence orbitals (often an ``active'' space, i.e. all valence configurations), but it has also seen a resurgence of interest in the form of selected configuration interaction, where $|\Phi_I\rangle$ is chosen in a problem specific manner to sparsely span the Hilbert space, and the number of configurations (in a small molecule) is treated as a convergence parameter, to converge to the exact result~\cite{sharma2017semistochastic,loos2020performance}. The lack of extensivity limits the application of configuration interaction to small problems.

\subsubsection{Perturbation theory and diagrammatic methods } 

\begin{table}\centering
\begin{tabular}{c|c}
\hline\hline
      Weak and strong & Weak (unless used with other methods) \\
       Systematically improvable & Yes (if converges) \\
       Extensive energy & Yes \\
       Cost scaling & Poly.  \\
       Prefactors & Very low \\
       Lattice models and ab initio & Both \\
     \hline\hline
\end{tabular}
\caption{Characteristics of perturbation theory and diagrammatic methods.
\label{tab:pt}}
\end{table}

 Another class of techniques is based on perturbation theory. 
In its simplest form, one evaluates the Taylor expansion (to finite order) of the ground-state energy $E(\lambda)$ corresponding to the Hamiltonian $H(\lambda) = F + \lambda W$, with $E^{(n)}$ being the $n$th order Taylor coefficient. In a more sophisticated (diagrammatic) form, we first write the energy as 
\begin{align}
    E=  \frac{\langle \Phi_0 | H \mathcal{T} e^{-\int_0^\beta d\tau V_I(\tau)}|\Phi_0\rangle}{\langle \Phi_0 | \mathcal{T} e^{-\int_0^\beta d\tau V_I(\tau)} |\Phi_0\rangle} = 
     {\langle \Phi_0 | H \mathcal{T} e^{-\int_0^\beta d\tau V_I(\tau)}|\Phi_0\rangle}_c
\end{align}
where 
$V_I$ is in the imaginary time interaction picture of $F$, $\beta$ is an imaginary time taken in the limit $\beta \to \infty$, and $\mathcal{T}$ is the time-ordering operator.
The diagrammatic structure arises by taking advantage of the simple ground-state for $\lambda=0$, i.e. $|\Phi_0\rangle$ is a Slater determinant, to recognize that the contributions at each order in perturbation theory to $E$ can be evaluated using Wick's theorem and visualized as a sum of diagrams. The second equality, and the subscript $c$ reflects the fact that only graphically fully connected diagrams survive the cancellation between the numerator and denominator.

Perturbation theory has a number of nice properties as illustrated in Table~\ref{tab:pt}. For example, it leads to an extensive energy (because $E(\lambda)$ is extensive, so its derivatives are extensive).  
As the sum of the zeroth and first-order energies is just the Hartree-Fock energy, the lowest non-trivial order is the second-order correction to the energy, conventionally denoted MP2 (Moller-Plesset second order perturbation theory). This is a fast and often qualitatively accurate correction to the Hartree-Fock results in systems where the Hartree-Fock energy gap is large (as is often true in small molecules)~\cite{cremer2011moller}. Applying perturbation theory to higher order rapidly becomes expensive, and need not lead to a better result because the theory usually diverges~\cite{olsen2000divergence}. Higher order perturbation theory contributions therefore usually come from one or more iterative techniques to resum certain classes of connected diagrams: these methods yield partial contributions to each order of perturbation theory up to infinite order~\cite{martin2016interacting}.

\subsubsection{Coupled cluster theory }

\begin{table}\centering
\begin{tabular}{c|c}
\hline\hline
      Weak and strong & Weak to medium \\
       Systematically improvable & Yes \\
       Extensive energy & Yes \\
       Cost scaling & Poly.  \\
       Prefactors & Low \\
       Lattice models and ab initio & Both \\
     \hline\hline
\end{tabular}
\caption{Characteristics of coupled cluster methods.
\label{tab:cc}}
\end{table}

The Hartree-Fock state can be viewed as generating an optimized Slater determinant in terms of another Slater determinant through the Thouless expression
\begin{align}
  |\Phi_\text{HF}\rangle = e^{\sum_{pq} A_{pq} c^\dag_p c_q}|\Phi_0\rangle
  \end{align}
The coupled cluster ansatz can be thought of as a generalization of the exponent, leading to the form
\begin{align}
  |\Psi\rangle = e^{T}|\Phi_0\rangle = e^{\sum_{pq} A_{pq} c^\dag_p c_q + \sum_{pqrs} A_{pqrs} c^\dag_p c^\dag_q c_r c_s + \ldots}|\Phi_0\rangle \label{eq:ccansatz}
  \end{align}
where $T$ is known as the cluster operator. $T$ is  expanded as a sum of terms in a many-body expansion where each term creates an individual fluctuation away from the $|\Phi_0\rangle$, but the exponentiation of $T$ means that $|\Psi\rangle$ contains global fluctuations parametrized as a product of fluctuations of smaller numbers (clusters) of particles.
In quantum chemistry applications, ${T}$ is further restricted to allow for the efficient evaluation of the needed computational expressions. For example, the excitation operator is restricted to be of the form
\begin{align}
  {T} = \sum_{ia} t_{ia} c^\dag_a c_i + \sum_{ijab} t_{ijab} c^\dag_a c^\dag_b c_i c_j + \ldots \label{eq:ccsd}
\end{align}
where $i,j$ label orbitals that are occupied in $|\Phi_0\rangle$ and $a, b$ label orbitals that are not.

Coupled cluster theory, for some low-order truncation of ${T}$ (usually to the first two terms in Eq.~\ref{eq:ccsd}, denoted singles and doubles, and with an approximate correction for the next term, the triples, known as CCSD(T)) is the most widely used many-electron wavefunction method in quantum chemistry~\cite{shavitt2009many}.
This is because it has good formal and practical properties and provides a good balance between cost and accuracy (see Table~\ref{tab:cc}); for example, when truncated to the singles and doubles level, the coupled cluster ansatz it is still exact for any problem that breaks down into sets of independent two-electron problems. 
In practice, for problems that are not too strongly correlated, CCSD(T) gives results for ground-state energy differences approaching ``chemical accuracy'' and, even for more strongly correlated problems, assuming an appropriate Slater determinant starting point can be found (see Sec.~\ref{sec:strongcorr}) it provides a reasonably accurate treatment of the quantum fluctuations around such a determinant. The main drawbacks of coupled cluster are that its costs grows exponentially with the truncation order, and thus it can become impractically expensive when  one does not have a good single Slater determinant starting point. (It can fail also in more significant ways that have not been fully analyzed, such as having no real valued solutions at a given  truncation order~\cite{paldus1984coupled}, and also the solution conditions can be hard to converge). These weaknesses manifest even if only a small number of Slater determinants are required in the multireference treatment if the participating determinants are separated by large number of excitations, perhaps as few as two. As a practical method, therefore, it does not  extend to the full set of problems covered by the simplifications discussed in the context of strong correlation in Sec.~\ref{sec:strongcorr}.

\subsubsection{Matrix product states, density matrix renormalization group, and tensor networks }

\begin{table}\centering
\begin{tabular}{c|c}
\hline\hline
      Weak and strong & Both; less efficient than other methods for weak \\
       Systematically improvable & Yes \\
       Extensive energy & Yes \\
       Cost scaling & Poly. for fixed $D$ \\
       Prefactors & High (DMRG), very high (TN)  \\
       Lattice models and ab initio & Both for MPS/DMRG, lattice for TN \\
     \hline\hline
\end{tabular}
\caption{Characteristics of matrix product state (MPS), density matrix renormalization group (DMRG), and tensor network (TN) methods.
\label{tab:tns}}
\end{table}

Matrix product states (MPS)~\cite{verstraete2008matrix} are a class of wavefunctions that are defined by a system-size independent amount of bi-partite entanglement. Writing $|\Psi\rangle$ in the occupation representation of some single particle basis, with $K$ orbitals,
\begin{align}
|\Psi\rangle = \sum_{ \{ n\} } \Psi^{n_1 \ldots n_K} |n_1 n_2 \ldots n_K\rangle
\end{align}
a matrix product state factorizes the amplitudes as a matrix product
\begin{align}
  \Psi^{n_1 \ldots n_K}  = \sum_{ \{ i\}} A^{n_1}_{i_1} A^{n_2}_{i_1 i_2} \ldots A^{n_K}_{i_{K}} \label{eq:mps}
  \end{align}
where the first and last matrices are vectors (so that the product is a scalar). For a given dimension of the ``bond'' indices $i_1 \ldots i_K$, $\mathrm{dim}({i}) = D$, the MPS can capture at most entanglement entropy of $\log_2 D$ between any left/right cut of the system between orbitals $i_1 \ldots i_l$, $i_{l+1} \ldots i_K$.

The structure of the MPS is quite different from that of the approximations so far discussed, which are based on limiting the complexity of excitations relative to a given Slater determinant $|\Phi_0\rangle$, such as the Hartree-Fock reference~\cite{chan2004state}. Assuming the occupancy basis in Eq.~\ref{eq:mps} is the Hartree-Fock orbital basis, the Hartree-Fock reference enters as just one of the possible occupancies, on much the same footing as the others. On the other hand, the flexibility of the amplitudes is severely limited as they must satisfy the near-product structure, and there is an inherent one-dimensional nature to the ansatz: more correlations can be captured between orbitals with indices $i, j$ if $|i-j|$ is small. In other words, MPS provide a parametrization  of quantum states that efficiently enforces the correct structure of locality in one dimension. In model systems, it can be proved that the ground-states of gapped local Hamiltonians in 1D have an efficient representation as a matrix product state~\cite{hastings2007area}. This means that the bond-dimension needed to represent the state to some accuracy is independent of system size.

In practice, however, MPS and the density matrix renormalization group (DMRG) algorithm that provides a practical way to variationally optimize MPS~\cite{white1993density,verstraete2023density}, have found applications significantly outside of pure 1D model Hamiltonians~\cite{stoudenmire2012studying}. The ansatz is always exact as $D$ is increased, and this, coupled with the efficient formulation in terms of matrix multiplications, and the lack of bias towards a Hartree-Fock (or other mean-field) reference, means that it is a good replacement for full configuration interaction/exact diagonalization for problem sizes that are too large for such methods and when there is not another simpler alternative ansatz. In quantum chemistry, it was one of the first widely used techniques for molecules with a large number of strongly correlated electrons~\cite{white1999ab,mitrushenkov2001quantum,chan2002highly,legeza2003controlling}, and remains an important  technique for challenging problems today~\cite{larsson2022chromium,baiardi2020density}. 

Tensor networks refer to the natural generalization of the matrix product state beyond the 1D entanglement structure~\cite{orus2019tensor}. Different kinds of tensor networks exist and some, such as projected entangled pair states, are now widely used with model Hamiltonians in two dimensions where they are a powerful alternative to MPS/DMRG. However, the algorithms that have been developed to work with them are currently only practical for model Hamiltonians; the overhead of more complicated long-range interactions is a formidable challenge~\cite{o2018efficient}. Extending these techniques to ab initio Hamiltonians remains an open scientific problem.

\subsubsection{Variational and Projector Monte Carlo }

\begin{table}\centering
\begin{tabular}{c|c}
\hline\hline
      Weak and strong & Both, may require special trial \\
       Systematically improvable & Yes (if trial is) \\
       Extensive energy & Yes (if trial is) \\
       Cost scaling & Poly. (if trial is)  \\
       Prefactors & High to very high  \\
       Lattice models and ab initio & Both  \\
     \hline\hline
\end{tabular}
\caption{Characteristics of quantum Monte Carlo methods used in quantum chemistry.
\label{tab:qmc}}
\end{table}

The methods mentioned so far are most commonly used in a deterministic setting. Another class of approximate methods can be formulated based on stochastic algorithms: these are known as quantum Monte Carlo algorithms~\cite{becca2017quantum}. There are two main families of quantum Monte Carlo algorithms used in quantum chemistry. The simplest is the variational Monte Carlo algorithm. This is based on sampling the amplitude $\Psi(n) \equiv \langle n|\Psi\rangle$ from the distribution $|\Psi(n)|^2$,  where here $|n\rangle$ is now considered to be the occupancy vector $|n_1n_2 \ldots n_K\rangle$. While one can find wavefunctions from which $\Psi(n)$ can be directly sampled, the Metropolis algorithm (assuming fast mixing) is all that is needed evaluate $\Psi(n)$. 
Given such samples, the energy can be evaluated as
\begin{align}
  E = \langle \Psi | H | \Psi\rangle = \sum_{nn'} |\Psi(n)|^2 \langle n|H|n'\rangle \frac{\Psi(n')}{\Psi(n)} 
  \end{align}

The advantage of variational Monte Carlo is that very general functional forms can be used: it is often easier to find functions for which $\Psi(n)$ can be efficiently evaluated, than to find ones for which $\langle \Psi|\Psi\rangle$ can be evaluated. (The first is analogous to evaluating a high-dimensional function value, while the second is analogous to evaluating a high-dimensional integral). In recent years, the rapid development of neural networks as general function approximators has led to the use of neural network parametrizations of $\Psi(n)$, which have been termed neural quantum states~\cite{luo2019backflow,hermann2020deep,pfau2020ab,hermann2023ab}. Note that the amplitudes can be written in the antisymmetric space of Slater determinants, or simply in the space of product functions: in the latter case, care must be taken to ensure that $\Psi(n)$ is antisymmetric with respect to particle intercharge.

Usually, evaluating a sample $\Psi(n)$ in variational Monte Carlo  scales better than the cost to deterministically evaluate the energy using another approximate method. Common variants exhibit the same sample scaling as mean-field theory. On the other hand, one needs to have a sufficient number of samples to control the stochastic error, and in large systems with many parameters, one needs to optimize them in the presence of the stochastic error. Since stochastic errors in different systems do not generally cancel, it is often necessary to compute energies to a fixed precision, rather than a fixed precision per particle. For a system size $L$, this adds an additional $L^2$ scaling to the computational cost.

One special aspect of quantum Monte Carlo methods, including variational Monte Carlo, is that it is the only method so far discussed that can work in a continuous spaces (e.g. by specifying a $N$-particle basis state by $3N$ continuous numbers, the list of positions of the electrons)~\cite{kent2020qmcpack}. Thus quantum Monte Carlo methods need not use a single particle basis.

The second main family of ground-state quantum Monte Carlo methods are projector quantum Monte Carlo methods~\cite{becca2017quantum,kent2020qmcpack,motta2018ab}. These are based on an implicit representation of the ground-state wavefunction as $\lim_{\beta \to \infty} \exp(-\beta H)|\Phi_0\rangle$, with a stochastic representation of $e^{-\beta H}$. The general idea is to write
\begin{align}
  e^{-\beta H} |\Phi_0\rangle  = e^{-\epsilon H} e^{-\epsilon H} \ldots e^{-\epsilon H} |\Phi_0\rangle
  = \sum_{\mu_1 \ldots \mu_T} (p_{\mu_1} O_{\mu_1})  (p_{\mu 2} O_{\mu_2}) \ldots (p_{\mu_T} O_{\mu_T} )|\Phi_0\rangle
\end{align}
where $p_{\mu_i}$ is a probability distribution with $\sum_\mu p_\mu O_\mu = e^{-\epsilon H}$, and the operators  $O_{\mu}$ are chosen to have the property that when acting on a Slater determinant they produce another (unnormalized) Slater determinant. Each Monte Carlo sample of the indices $\vec{\mu} \equiv \mu_1 \ldots \mu_T$ yields an elementwise choice of operators $O_{\mu_1} \ldots O_{\mu_T}$ that may be thought of  as a path through imaginary time, and produces a representation of the final wavefunction as 
\begin{align}
  |\Psi\rangle = \sum_{\vec{\mu}} w_\mu |\Phi_{\vec{\mu}}\rangle
\end{align}
where $\Phi_{\vec{\mu}}$ need not be unique (i.e. different paths can give the same determinant, or determinants which are not orthogonal). 
However, because the weights $w_\mu$ can have either sign (or even be complex, depending on the decomposition into operators $O_\mu$) and in general $w_\mu$ spans an exponentially large range (coming from the multiplication of many numbers), the final representation of $|\Psi\rangle$ emerges from the cancellation of large weights. This is the fermion sign problem.

The sign problem is generally removed by introducing a trial wavefunction which contains information on the sign; the way in which the trial wavefunction is used to remove the sign depends on the type of projector Monte Carlo, but in all cases the outcome of the walk no longer samples the exact wavefunction but some approximation to it. However, all methods employing a trial wavefunction converge to the exact result as the trial wavefunction improves.

Projector Monte Carlo avoids the need to optimize a large number of variables as in variational Monte Carlo. When a simple trial state is sufficient it thus inherits some of the favourable scaling of quantum Monte Carlo methods, and it also has the flexibility to incorporate complicated trial states which capture features of strong correlation, although the need to beat down the stochastic error with a large number of samples remains.

\subsection{Summary}

We briefly summarize our survey of classical heuristics  in quantum chemistry and the intuition behind them. The main intuition is provided by mean-field theory (most commonly density functional theory) where the state of interest is viewed as an essentially classical (non-entangled) state in an appropriate optimized basis. This intuition serves well for many problems. For example, even in cases of strong correlation when a specific low-energy mean-field reference is inappropriate, the electronic structure may still be understandable in terms of the low-lying manifold of different mean-field solutions, or a linear combination of a small number of them to reflect fluctuations from a locally strong perturbation, or (in a material) a polynomial number of connected fluctuations. In cases where even the latter picture breaks down, locality remains a useful simplification and is the basis for more advanced heuristics. Although one can imagine quantum ground-states where these intuitions do not apply (for slightly more discussion, see Sec.~\ref{sec:eqa}) in the author's experience, the above three scenarios cover the vast majority of problems currently studied in quantum chemistry. 

In the numerical simulations of electron correlation beyond mean-field theory, there are a wide variety of approximate methods that can be applied. Surveying the tables of characteristics of methods discussed, however, we can identify some immediate  gaps, in particular in the treatment of systems with both a large amount of strong  and weak correlation. This gap may not be one of principle: for example, there is no reason a priori to expect that tensor networks cannot be used in a more continuum like (complete basis) description~\cite{haghshenas2021numerical}, and perhaps, truncated coupled cluster methods with sufficiently high levels of truncation can be used to describe the strongly correlated problems of interest in chemistry; but such methods do not yet exist today, and will certainly need new ideas if they are to be practical in the future. 
We return to this point in Sec.~\ref{sec:classicalrefinement}.

The fact that there is a gap in capabilities does not conflict with the possibility of achieving an intuitive understanding of correlated electron systems: even for problems where mean-field methods are an appropriate description of the electronic structure (for example, in the ligand-binding example in Sec.~\ref{sec:openproblems}) there can still be a formidable computational gap to making quantitatively precise predictions. At the same time, some of these gaps (particularly in the systems with both substantial strong and weak correlations~\cite{larsson2022chromium}) seem to arise because the current set of classical heuristics do not fully embody the intuitive simplifications that exist in physical systems. This suggests that there are new conceptual directions to explore.

The methods above as usually employed are polynomial cost heuristics. Aside from the tensor network methods, the techniques do not incorporate locality as an assumption from the outset. Variants of these approaches that further incorporate locality can be devised, and under the (often chemically reasonable) assumption of strong locality (see Sec.~\ref{sec:locality}),  
these reduce the cost scaling to linear in the system size. Although there is no formal theory of error associated with these heuristics, there is a lot of empirical understanding of when they work and when they do not. Thus the judicious application of these methods, sometimes in concert with each other, has helped establish the success of computational quantum chemistry.

\section{The complexity of quantum chemistry using classical heuristics}

In the prior sections we have briefly described the intellectual framework of quantum chemistry in the context of the electronic structure problem and the intuition that can be established about the low-energy states. In the current section, we consider the consequences of the conjectured behaviour of chemical matter on the classical complexity of quantum chemistry. For example, given all these assumptions, should we consider the simulation of quantum chemistry using classical heuristics to be easy or hard? 


Heuristics are usually run with polynomial cost, but without a detailed understanding of the error, but to discuss the complexity one needs to understand the associated error as well.
In other words, we seek to obtain the cost $C$ of running a heuristic to some desired error $\epsilon$, for a given system size $L$, the function $C = f(L, \epsilon)$.

\subsection{An aside on absolute and relative errors and chemical accuracy}

We first discuss what kind of error we usually desire. Today the term chemical accuracy often means an error in the total electronic energy of 1~kcal/mol (about $kT$ at room temperature, the ambient chemical temperature). But the significance of $kT$ is related to energy differences, not the total energy (which after all contains arbitrary constants, for example, the nuclear-nuclear energy at a fixed geometry). Furthermore, if one considers a large system (or even a system in the thermodynamic limit, where the energy diverges) it does not seem sensible to compute the total energy to 1~kcal/mol.

The term chemical accuracy was originally coined in the context of chemical energy differences~\cite{pople1999nobel}. When applied to total energies, one could argue that a more reasonable definition of chemical accuracy is in terms of the relative error $\bar{\epsilon} = \epsilon / N$ (and up to a multiplicative factor, $\epsilon/L$). Ideally this applies to the energy of the valence electrons (where chemistry takes place) but in practice it is difficult to separate out such an energy component. Leaving that aside, the relative error is a more appropriate metric for chemical reactions, as these involve local changes, and it is also consistent with assuming equivalence of statistical mechanical ensembles in the thermodynamic limit.

Nonetheless, almost all numerical calculations compute the total energy of a problem rather than the relative energies directly: relative energies are obtained as the difference of the total energies. We might then wonder whether computations should target a relative error or an absolute error in the total energy? 

Because of the observed locality of chemical matter, it turns out that many heuristics for total energy computation can yield useful answers even if one only targets the relative error rather than the total error. For example, consider the basis set error in a chemical reaction. The basis set error in the total energy is a global quantity which (for a given basis resolution) simply increases with the size of a system. However, in a chemical reaction, where changes are only occurring locally, then the energies of the regions far away cancel out between, say, the reactants, and the products. Under the setting of strong locality, most of the classical approximations with the ``extensivity'' property in Sec.~\ref{sec:survey} produce an energy error which is approximately additive across the atoms in the simulation. Then only the local error matters in modeling a chemical reaction.

The main exception to the above is stochastic errors, which (without using special types of sampling) are uncorrelated between different calculations. Uncorrelated stochastic errors in calculations where an energy difference is to be taken must therefore be converged to a given absolute error, in order to obtain the same order of error in the energy differences.

\subsection{The role of asymptotic analysis}

\label{sec:asymptotic}
We have previously expressed intuitions that argue for a finite length scale of relevant quantum simulations (be it from the finite problem size, an effective symmetry breaking past some size, or arising from the intrinsic locality of interactions). We have also argued for a finite desired accuracy due to both limited experimental control and the relevant conditions, roughly $O(kT)$ at ambient temperature. Since complexity usually refers to asymptotic behaviour with respect to $L$ and $\epsilon$, how relevant is asymptotic analysis?

Although the limit of $L \to  \infty, \epsilon \to 0$ is not chemically very relevant, it is still useful to understand the scaling of costs and errors up to the relevant cutoff scales, which we might denote $L_c$ and $\epsilon_c$. Of course it is difficult to agree exactly what these cutoffs are, especially for $L_c$. To limit the scope of discussion, we can consider only the case of modeling electron correlation in ab initio calculations. Then, one pragmatic perspective is that we would like to model correlated electron effects for the same systems that one can routinely perform mean-field calculations for, say $\sim O(1000)$ atoms and $O(10000)$ electrons, or a linear dimension of $O(10)$ atoms in a three-dimensional problem. (Under chemical conditions, the minimum spacing of atoms is $O(1)$ atomic units, thus this translates into a length scale as well).



If a system exhibits strong locality, then one would expect that on scales larger than the associated $L_c$, the cost to simulate the system is $O(L)$. If there were no physical structure to the problem before the cutoff scale is reached, then up to size $L_c$, the observed $C(L) \sim e^{O(L)}$; subsequently across a class of problems where $L_c$ can be tuned, the cost scaling would be expected to grow like $e^{O(L_c)}$. But in the author's experience, this is too pessimistic, as the various physical structures of correlation impose themselves before $L_c$ is reached. For example, in the weak correlation limit, a truncated coupled cluster calculation may be sufficient, with $O(\mathrm{poly}(L))$ scaling, before systems sizes of $O(L_c)$, while e.g. in tensor networks, as one tunes correlation functions towards a critical system (thereby increasing $L_c$), the ground-state entanglement does not change from an area law entanglement to a volume-law entanglement at the critical point, but only acquires logarithmic corrections. Thus for a given error, we expect to see a smooth crossover between the functional form of $C(L)$ at small $L$ and for $L > L_c$.

\subsection{Two kinds of error from classical heuristics}

We now suggest that it is useful to think about classical heuristics as exhibiting two kinds of errors. The first is a ``reference' error. In the language of quantum algorithms discussed later, we might view this as related to state preparation, as we will discuss. The second is a refinement error.

We have already described the various kinds of electronic structure observed in the ground- and low-lying excited states in Sec.~\ref{sec:survey}. These different qualitative starting points might be thought of as the references. Thus the reference error is associated with constructing an appropriate choice of starting point. The refinement error is the error associated with applying a heuristic after that starting point has been constructed.

In a molecule, a prototypical example of these two types of errors can be seen in the case of stretching a chemical bond. Consider the case of using the coupled cluster heuristic. This requires a Slater determinant starting point (such as the restricted Hartree-Fock reference) which we observe is usually qualitatively correct near the equilibrium geometry but becomes increasingly poor as the bond is stretched. The coupled cluster calculation (truncated to singles and doubles) is then very accurate at the equilibrium geometry, and the remaining error from the exact calculation (in this basis) can be viewed as the refinement error. However, the same coupled cluster calculation becomes poor as the bond is stretched. 

Because the coupled cluster method is systematically improvable, we could in principle improve the result at long bond lengths by increasing the truncation level at additional cost.
But, in the stretched region, one finds that another type of Hartree-Fock solution can be found (the unrestricted Hartree-Fock solution) and this qualitatively describes the energetics of the bond dissociation. Introducing the coupled cluster ansatz on top of the unrestricted Hartree-Fock solution then leads to much more accurate energies at stretched geometries (see Fig.~\ref{fig:ccbond}). 

Alternatively, we could construct a reference consisting of the important configurations around the Hartree-Fock reference. If we include all the valence excitations, associated with the triple bond breaking, we obtain the complete active space self-consistent field (CASSCF)) reference~\cite{helgaker2013molecular}. This reference state contains (but is not limited to) all Slater determinants that become degenerate or nearly degenerate at long distance.
We see that its energy is now qualitatively correct both at short and long bond lengths. Refining this now by including singles and doubles type fluctuations on top of it (the MRCISD~\cite{helgaker2013molecular} curve) yields an almost exact result.

\begin{figure}[!htbp]
  \includegraphics[width=275px]{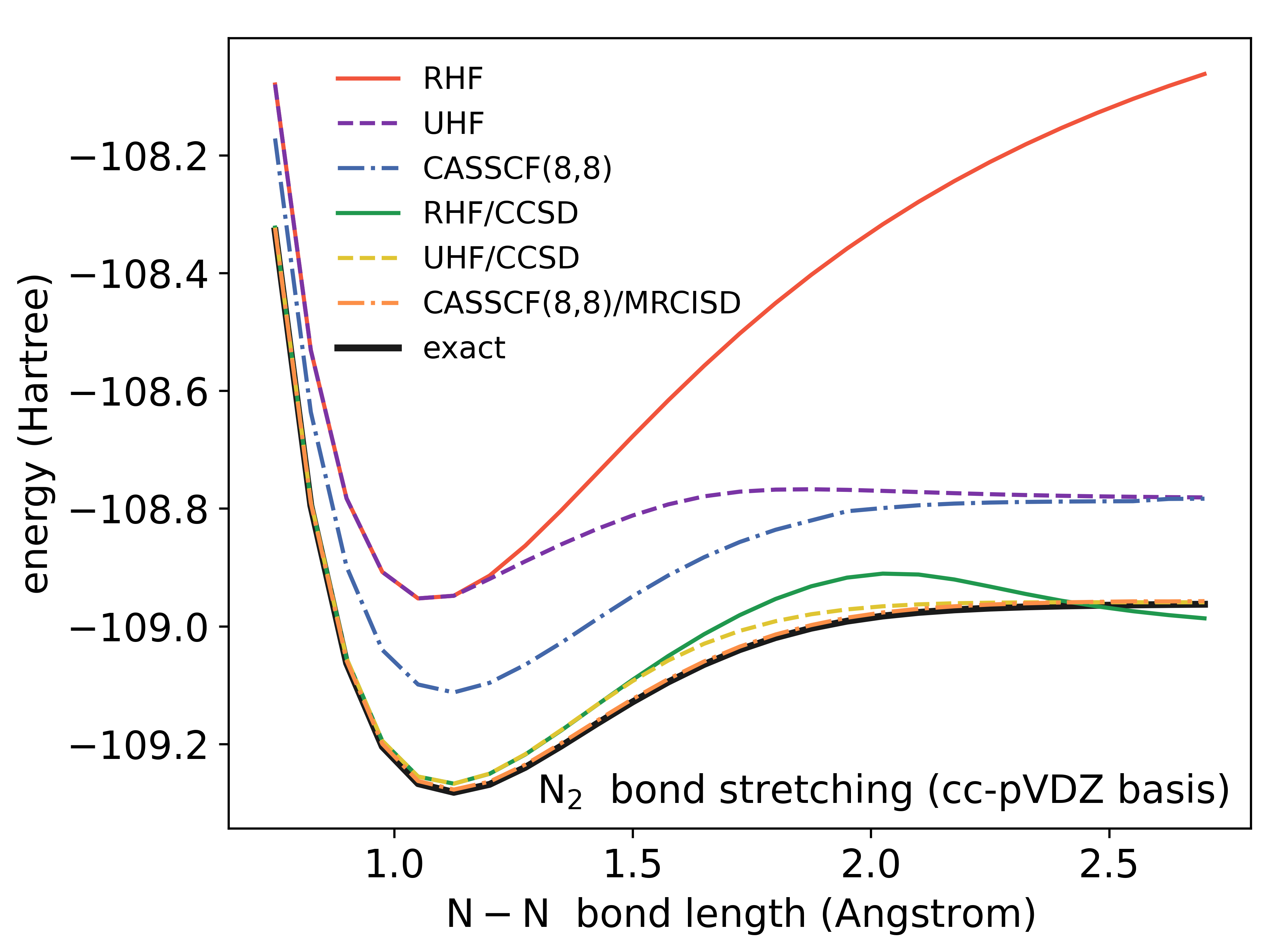}
  \caption{The dissociation curve of the N${}_2$ molecule, computed in the cc-pVDZ basis. 3 references are shown: the restricted Hartree-Fock (RHF) and unrestricted Hartree-Fock (UHF) references, which are both Slater determinants, and the complete active space self-consistent field (CASSCF) reference, which includes all excitations of the valence electrons. On top of the Hartree-Fock references, we show results from the coupled cluster with singles and doubles (CCSD) ansatz, while on top of the CASSCF reference, we show results from the uncontracted multireference configuration interaction with singles and doubles (MRCISD) method.}
  \label{fig:ccbond}
\end{figure}

Thus we can see that it is useful to think of the error associated with the choice of reference as a distinct error from the error associated with refining it. In the coupled cluster truncation, this is because the cost to find another Hartree-Fock solution that is appropriate for longer bond lengths will be much less than the cost to improve the coupled cluster calculation starting from a poor choice of reference. 
In the case of using the CASSCF reference, it is the fact that we are making an appropriate choice of Slater determinants to linearly combine which leads to the qualitatively correct curve; the refinement on top of this then requires only low-order fluctuations.
In the materials setting, we can regard the idea of starting from the correct reference point as starting from a state that is in the correct quantum phase. 

Accepting that there are these two types of errors, we may view a classical heuristic algorithm as containing two steps: first, finding the appropriate reference state (state preparation) and second, preparing it and refining its energy. The cost function then takes the form
\begin{align}
  C = R_\text{state prep} \times C_\text{refinement}
  \end{align}
  Here $R_\text{state prep}$ refers to the number of different kinds of states that must be prepared to find the correct starting point (the number of times state preparation is repeated) -- this is not the same thing as the number of determinants in a multi-reference state that is a linear combination of determinants, as a multi-reference state is a single state in this accounting; but in the case of the coupled cluster calculations above where we tried both a restricted Hartree-Fock reference and an unrestricted Hartree-Fock reference, $R_\text{state prep}=2$. The actual cost of preparing the given state is contained in $C_\text{refinement}$, which contains this cost and that of the subsequent heuristic applied on top of it.
We now examine these two pieces in more detail.

\subsection{The difficulty of state preparation in classical heuristics}
\label{sec:classicalstateprep}

How difficult is it to write down a good starting point for a classical quantum chemistry heuristic? 

There are two logical possibilities. The first is that the starting point is intrinsically hard to describe classically  i.e. the quantum state does not even have a succinct classical description. (By classically succinct we mean here that the state can be stored with a polynomial amount of classical information and that its relevant properties extracted with a polynomial amount of classical computation).

If there is no such succinct description, that means that the state cannot be qualitatively expressed or interrogated in a faithful form without a beyond classical, i.e. quantum, computer. It would be some kind of black box for which, at least for certain properties, we could not write down a ``theory'' in the usual sense. Although such states can certainly be imagined (and manipulated, say, on a quantum computer) to the best of the author's knowledge, they have never been observed in the low energy states of chemical matter. Indeed, it is a remarkable feature of Nature that despite the vastness of the Hilbert space of a material in the thermodynamic limit, the number of ground-state phases observed is relatively small (e.g. insulators, metals, superconductors, various topological orders etc.) -- at least compared to the size of the Hilbert space. Furthermore, simple qualitative wavefunctions (and theories) have been formulated to describe the observed phases, such as Fermi liquid theory, Landau's theory of symmetry broken phases, the Bardeen-Cooper-Schrieffer wavefunction, the Laughlin wavefunction, and so on. As argued already in Sec.~\ref{sec:strongcorr}, we can even make an empirical claim that much of the chemical matter actively studied in chemistry can be thought of as qualitatively close to some mean-field state, perhaps with a broken symmetry, or with a polynomial number of connected fluctuations away from it. Thus, we conjecture that in chemical matter the starting point can be expressed in a classically succinct manner.

With the above assumption, the main remaining challenge comes from finding this classical starting point. One strategy is to search over the possibilities, i.e. enumerate the different starting points, apply a refinement (using one of the many classical heuristics) and see which yields the best answer (i.e. the lowest energy). Indeed, this is what is done in practice in particularly challenging problems. For example, in the study of the Fe-S clusters in nitrogenase, in practice one generates many different broken-symmetry mean-field solutions~\cite{sandala2011modeling,jiang2023protonation}, which each can then be refined using more sophisticated techniques, such as the density matrix renormalization group~\cite{li2019electronic}. Similarly, in condensed matter simulations of the Hubbard model, one applies different kinds of pinning fields and boundary conditions to ``prepare'' different ordered phases which are in close competition in the phase diagram~\cite{zheng2017stripe}. 

While we do not have a rigorous enumeration of the possible phases of matter, it seems reasonable to assume that in the worst case  (over systems of chemical relevance) we can search over a number of possibilities that is exponential in system size. This exponential complexity is realized in finding the ground-state of classical spin glasses, for example, where the complexity class for ground-state determination is \textsc{NP} hard. Given that we expect the starting point to be classically succinct, we can conjecture additionally that the state-preparation problem in chemical matter is then in fact \textsc{NP} hard (loosely, classically exponentially hard). This is a simplification of the \textsc{QMA} hardness (i.e. quantumly exponentially hard) that is expected for the general ground-state problem, and which is {realized in quantum spin glasses}. (In essence, this simplification is equivalent to stating the worst case quantum spin glass problems are not in the set of today's chemically relevant problems). 

Of course, all the above discussion also does not mean that we will always encounter exponential complexity of state preparation in practice: most chemical systems studied are not spin glasses of any type. In those cases where empirical inference tells us that we can prepare a simple mean-field state, or perhaps, a linear combination of such mean-field references e.g. to describe a bond breaking, $R_\text{state prep} = 1$.

\subsection{The cost of refinement}

\label{sec:classicalrefinement}

Given a reasonable starting point, how well do classical heuristics refine the result? In other words, given a desired error $\epsilon$ or relative error $\bar{\epsilon}$, what is the functional dependence of $C$ on the error? 
Despite the long history of quantum chemistry, unfortunately, this has not been much studied in  the community, and it constitutes an important open area of research. Note that while the author's opinion is that there is likely always a classically succinct starting point in chemical matter (as discussed above) this does not guarantee that we can efficiently refine the energy from it. For example, the error scaling may be impractically poor, such as $e^{O(1/\bar{\epsilon})}$, or there may simply be no heuristic to refine the error at all! Below we will summarize below some partial results to address this question from Ref.~\cite{expadvantage}, and we  present some new analysis as well.

In Ref.~\cite{expadvantage}, we first considered a set of molecules for which the Hartree-Fock reference is a good starting point, where the energy is further refined using coupled cluster theory. We can plot
the error of the coupled cluster theory as a function of the truncation order, and estimate the cost of the associated coupled cluster calculation (from the cost of the most expensive tensor contraction). 
(To be precise, this is not a perfect
accounting of costs as (i) it does not account for the number of such contraction terms or (ii) the number of iterations needed to solve the amplitude equations. To our knowledge, (i) has not previously been examined, but 
for the most common case where the number of virtual (unoccupied) orbitals is much larger than the number of occupied orbitals, the number of such terms is $O(1)$).



\begin{figure}[!htbp]
  \includegraphics[width=275px]{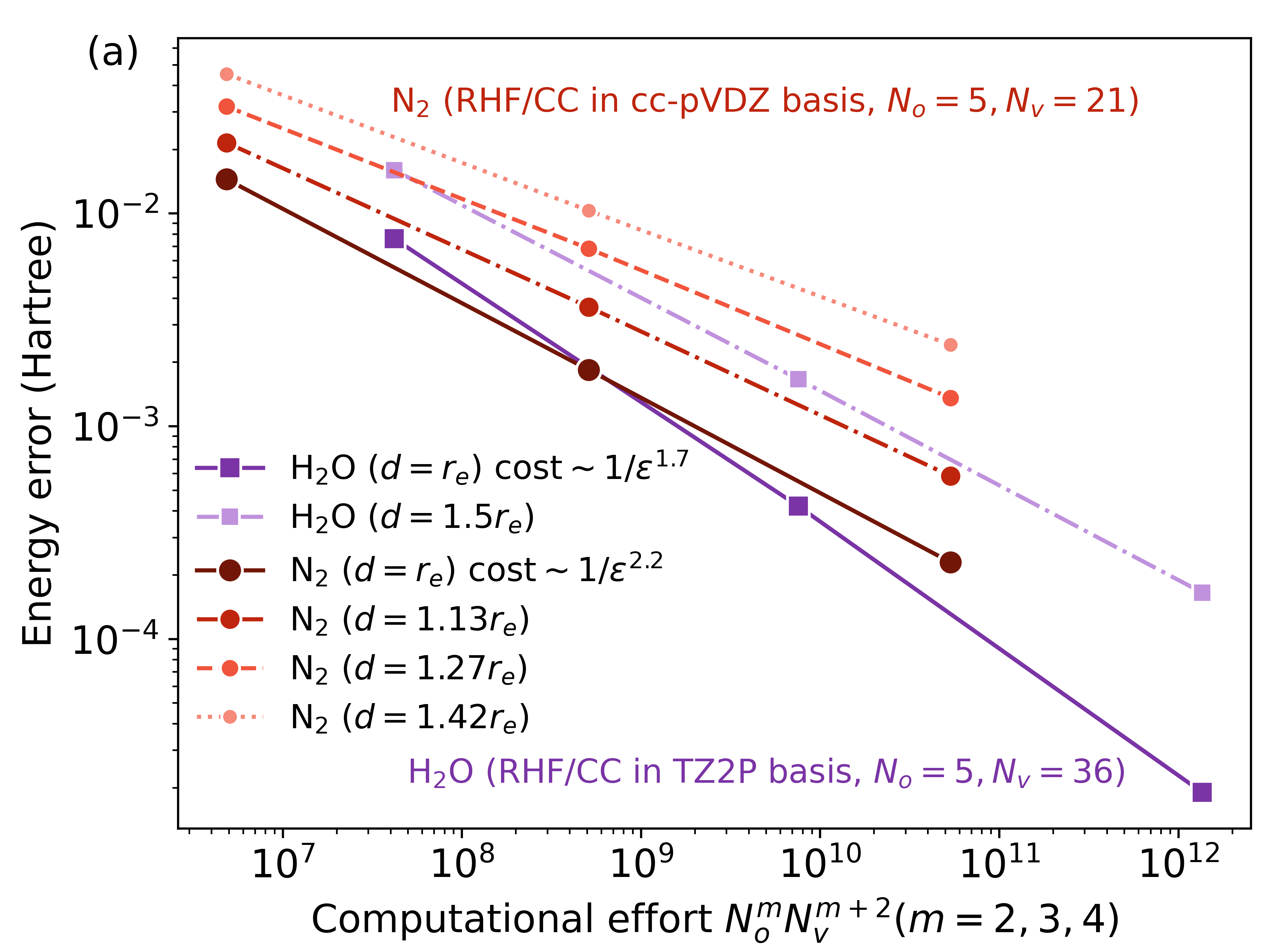}
  \includegraphics[width=275px]{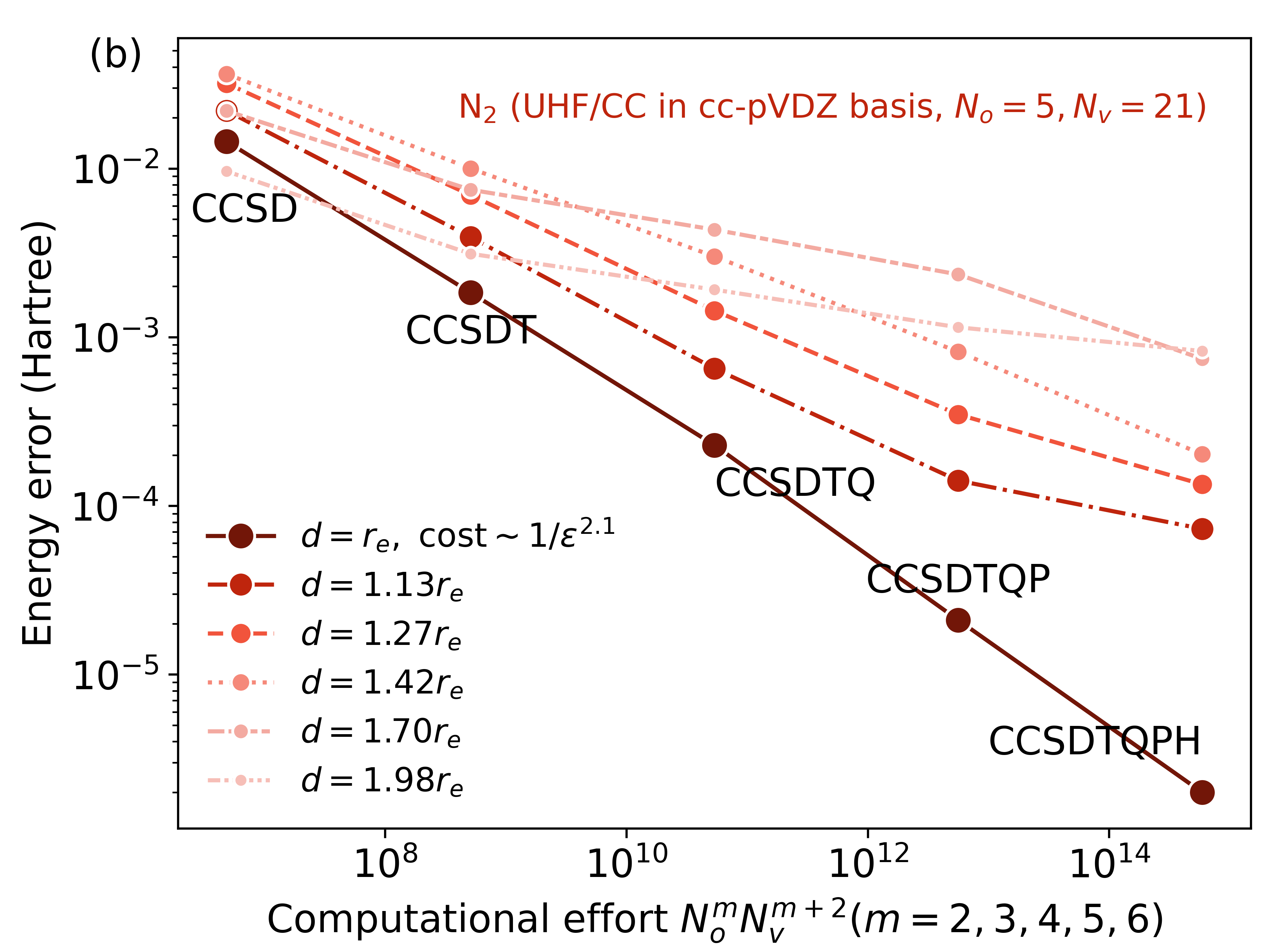}
    \caption{Energy error of nitrogen and water molecules at equilibrium $r_e$ and stretched (multiples of $r_e$) bond lengths, as a function of the level of coupled cluster truncation, against a computational cost metric. (a) and (b) show the data computed using RHF/CC and UHF/CC, respectively. Data taken from Refs. \cite{chan2004state} and \cite{chan2003exact}.}
  \label{fig:ccdata}
\end{figure}

In Fig.~\ref{fig:ccdata} we show data for the nitrogen molecule and the water molecule, at various bond lengths and (in the case of the nitrogen molecule) starting from restricted and unrestricted Hartree-Fock references. (The data for $\ce{N_2}$ starting from the unrestricted Hartree-Fock reference at the equilibrium geometry was previously discussed in Ref.~\cite{expadvantage}, but is here presented with a more precise accounting of computational cost).

The straight line fits for the $\ce{N_2}$ and $\ce{H_2O}$ molecules on the log-log plot of error (relative to the full configuration interaction result in the given basis) versus cost corresponds to a functional form of the cost of $\sim 1/\epsilon^\alpha$ (i.e. $O(\mathrm{poly}(1/\epsilon))$). Because of the exponential ansatz formulation of coupled cluster theory, for a gas of such non-interacting molecules, using the same parameters in the exponential operator yields the same relative error. Thus assuming this cost form, the total empirical cost of coupled cluster for a gas of these molecules is $O(\mathrm{poly}(L)\mathrm{poly}(1/\bar{\epsilon}))$ which translates to a cost of $O(\mathrm{poly}(L)\mathrm{poly}(1/\epsilon))$. Given the high quality of fit for the close to equilibrium geometries of the two molecules, we might conjecture that this is the cost scaling of coupled cluster theory for such so-called single reference problems. In Ref.~\cite{expadvantage}, we give further evidence, using local coupled cluster theory~\cite{liakos2015possible}, that indeed this is the appropriate scaling in large organic molecules.

An important feature of these error plots, however, is that the slope of the curves shows substantial system dependence, on the molecule, the bond length, and the starting reference that is chosen. At the equilibrium geometries, the slope is large, indicating a modest polynomial dependence on $1/{\epsilon}$ close to $O(1/{\epsilon}^2)$. However, as the bond lengths are stretched (to some multiple of $r_e$, the equilibrium lengths), and the problem is more and more multireference, the convergence becomes slower. For some (but not all) stretched bond lengths, the convergence is less systematic than at the equilibrium geometry. In table~\ref{tab:fit-r2} we show the quality of fit to both $\sim 1/\epsilon^\alpha$ and $\sim e^{\alpha/\epsilon}$ for the two molecules for the different bond lengths and starting references. We see that in all cases except one, the cost indeed fits $\sim 1/\epsilon^\alpha$ very well ($R^2 \sim 0.98$ or larger) and does not at all resemble the exponential inverse error cost form. However, at the longest stretched bond (1.98 $r_e$) for $\ce{N_2}$, the exponential inverse error form fits slightly better than the polynomial inverse error, and even in the polynomial fit, the error scaling is as high as $O(1/{\epsilon}^7)$). 

The highly stretched nitrogen geometry might then be taken as an example where the coupled cluster heuristic (over a relevant range of errors) does not efficient refine the energy. This is to be expected, as it is the nature of heuristics that they cannot work efficiently in all problems. But this does not rule out the use of a different heuristic for this situation which may be efficient. In Fig.~\ref{fig:dmrg} we show the error versus computational cost of DMRG for the nitrogen molecule at the same geometries. Unlike in the case of coupled cluster theory, the DMRG calculations refine the energy equally well at the different geometries and show $O(\mathrm{poly}(1/\epsilon)$ error scaling in every case (in fact the cost scaling appears to be better than $\mathrm{poly}(1/\epsilon)$, as expected from some theoretical arguments~\cite{chan2002highly}).

\begin{figure}[!htbp]
  \includegraphics[width=275px]{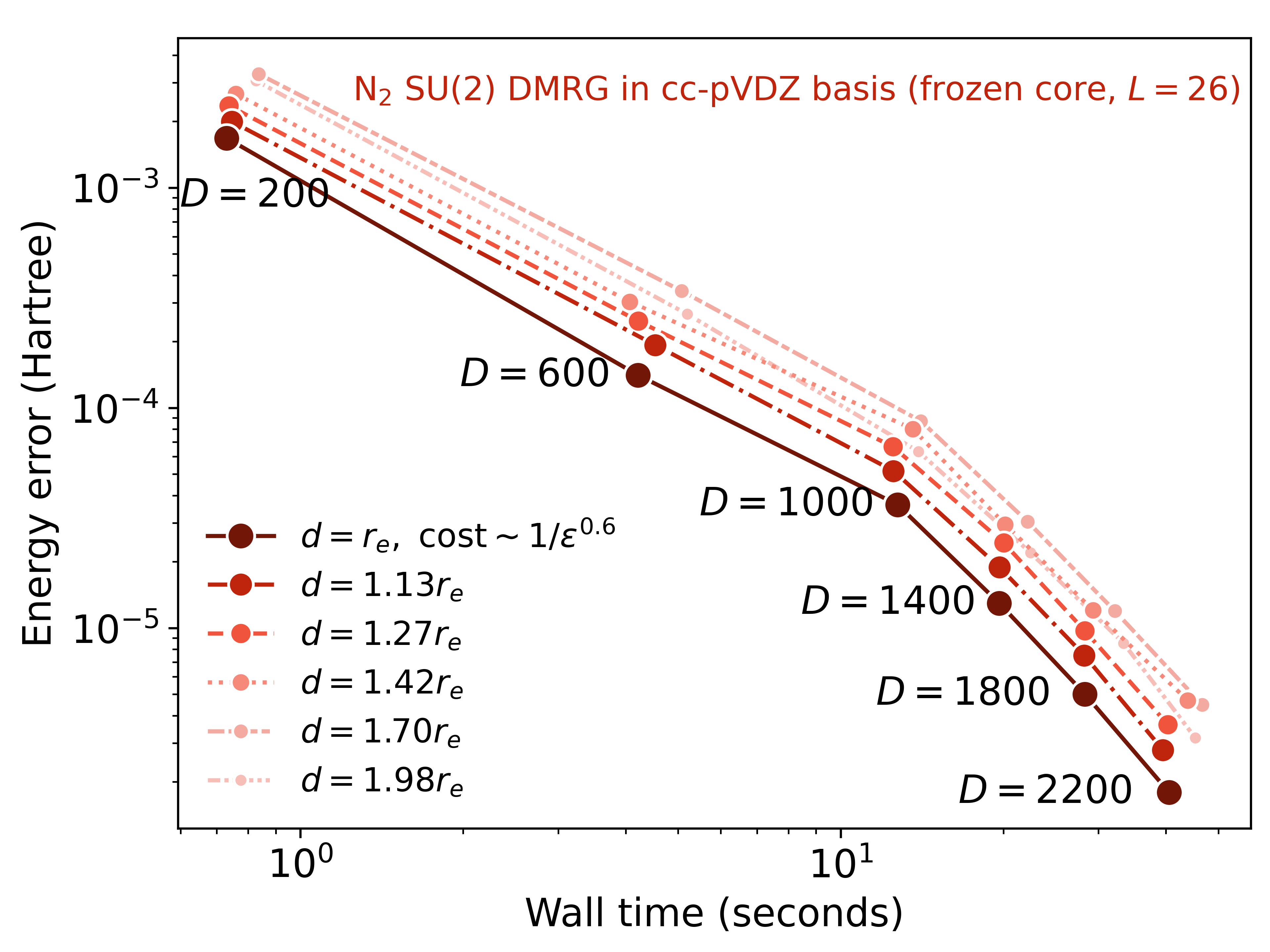}
    \caption{{Energy error of nitrogen at equilibrium $r_e$ and stretched (multiples of $r_e$) bond lengths computed using spin-adapted quantum chemistry DMRG, as a function of wall time cost per DMRG sweep for MPSs with different SU(2) bond dimensions $D$. Calculations use restricted Hartree-Fock orbitals ordered by point group irreducible representations.}}
  \label{fig:dmrg}
\end{figure}

\begin{table}\centering
\begin{tabular}{
    >{\centering\arraybackslash}p{1.0cm}>{\centering\arraybackslash}p{0.8cm}
    >{\centering\arraybackslash}p{1.0cm}>{\centering\arraybackslash}p{1.0cm}
    >{\centering\arraybackslash}p{1.0cm}>{\centering\arraybackslash}p{1.0cm}
    >{\centering\arraybackslash}p{1.0cm}>{\centering\arraybackslash}p{1.0cm}
}
    \hline\hline
      \multicolumn{2}{c}{Method} & \multicolumn{2}{c}{RHF/CC} & \multicolumn{2}{c}{UHF/CC} & \multicolumn{2}{c}{DMRG} \\
      System & $d$ & $R^2_{\mathrm{poly}}$& $R^2_{\mathrm{exp}}$
       & $R^2_{\mathrm{poly}}$& $R^2_{\mathrm{exp}}$
        & $R^2_{\mathrm{poly}}$& $R^2_{\mathrm{exp}}$\\
      \hline
      \multirow{2}{*}{$\mathrm{H_2O}$}
        &    $r_e$ & 0.9996 & 0.7820 &        &        &        &        \\
        & $1.5r_e$ & 1.0000 & 0.8168 &        &        &        &        \\
      \hline
      \multirow{6}{*}{$\mathrm{N_2}$}
        &    $r_e$ & 1.0000 & 0.8317 & 0.9988 & 0.5732 & 0.9743 & 0.4327 \\
        &$1.13r_e$ & 0.9999 & 0.8507 & 0.9775 & 0.8449 & 0.9713 & 0.4378 \\
        &$1.27r_e$ & 0.9998 & 0.8684 & 0.9925 & 0.7868 & 0.9714 & 0.4451 \\
        &$1.42r_e$ & 1.0000 & 0.8868 & 0.9993 & 0.6942 & 0.9732 & 0.4672 \\
        &$1.70r_e$ &        &        & 0.9815 & 0.7510 & 0.9721 & 0.4487 \\
        &$1.98r_e$ &        &        & 0.9436 & 0.9843 & 0.9750 & 0.4355 \\
    \hline\hline
\end{tabular}
\caption{{$R^2$ value of the quality of fit to $C \sim 1/\epsilon^\alpha$ and $C \sim e^{\alpha/\epsilon}$ for the coupled cluster and DMRG methods, computed for nitrogen and water molecules at different bond lengths $d$. See Figs. \ref{fig:ccdata} and \ref{fig:dmrg}.}
\label{tab:fit-r2}}
\end{table}

\begin{figure}[!htbp]
  \includegraphics[width=275px]{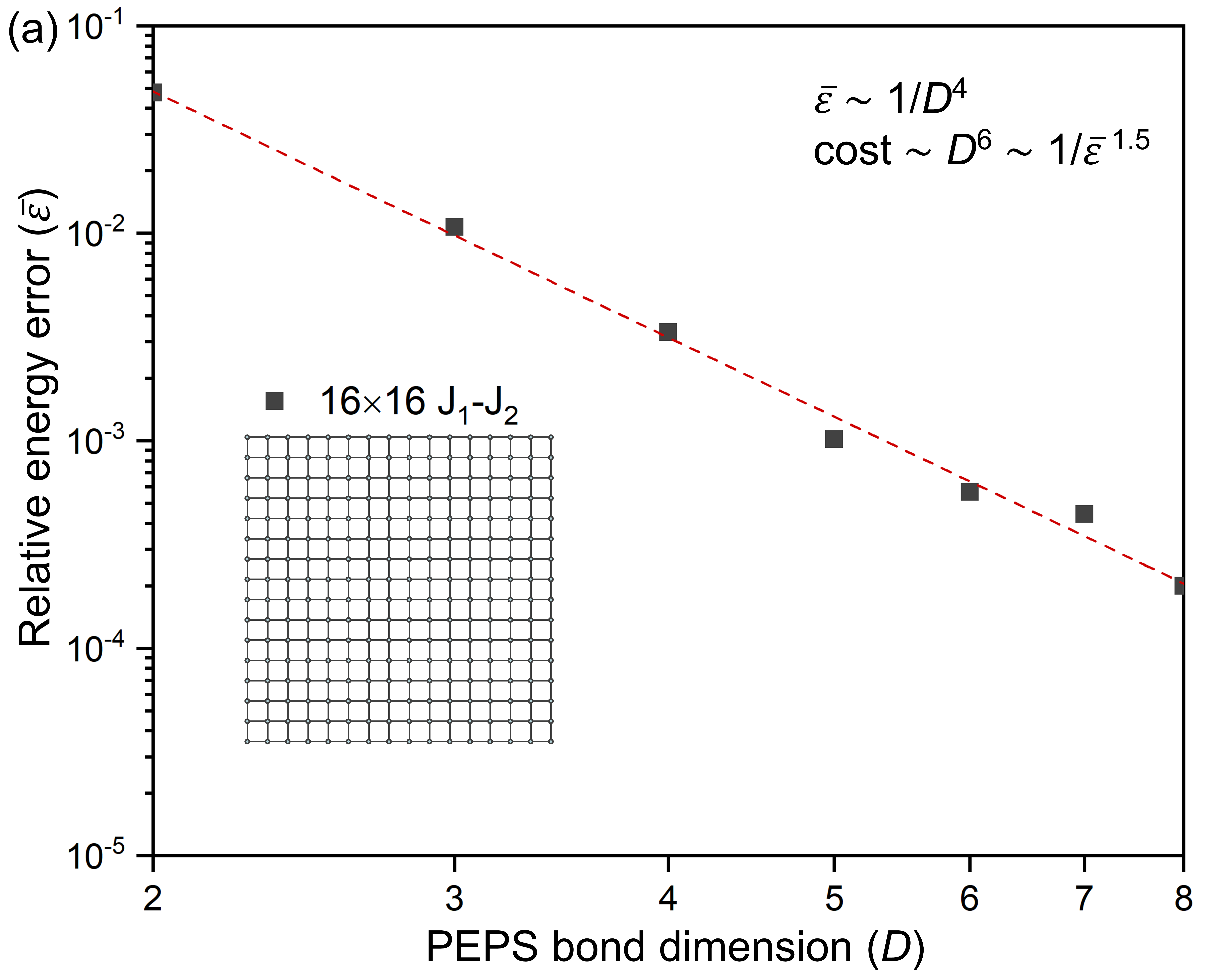}
  \includegraphics[width=275px]{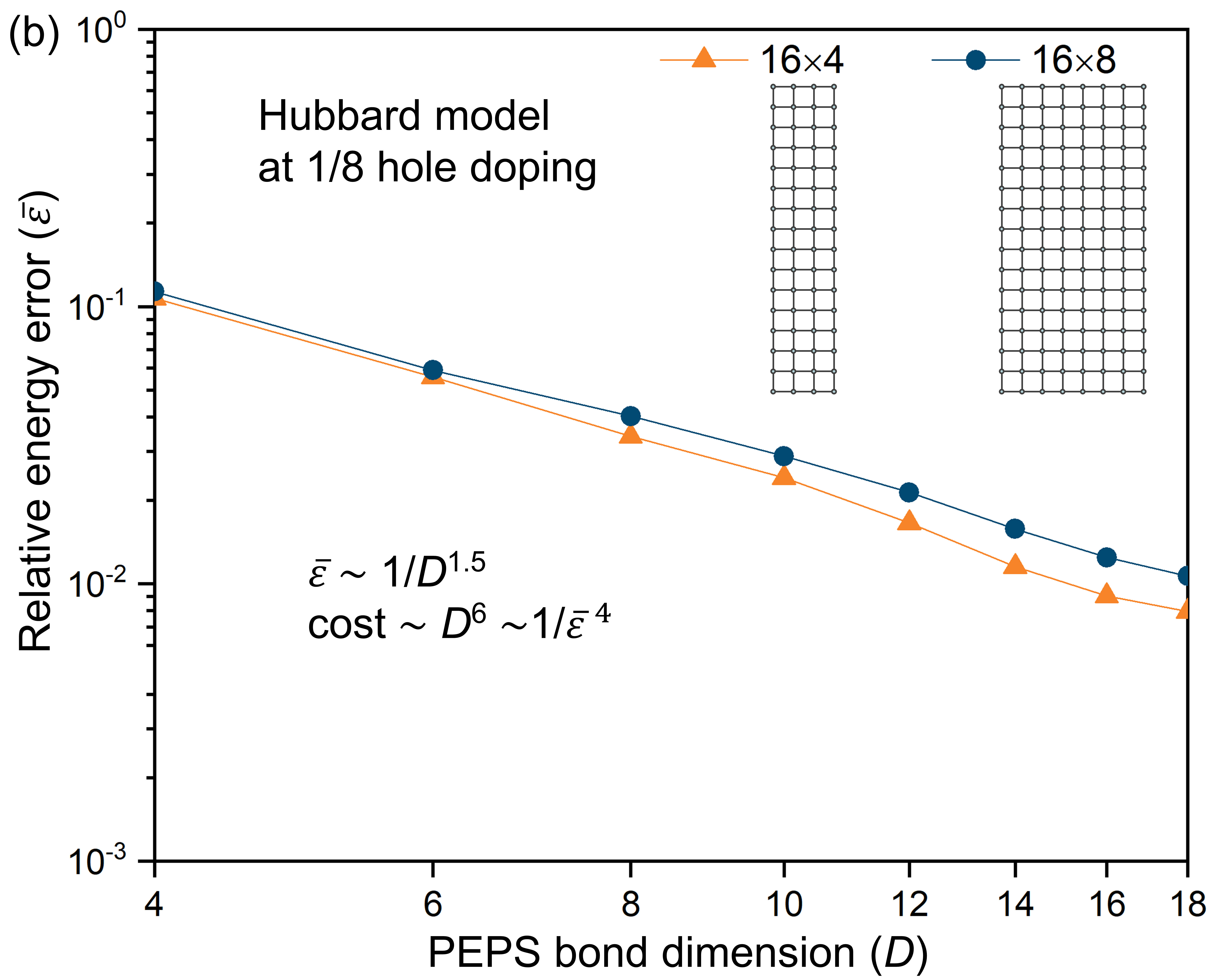}
    \caption{The log-log plot of energy accuracy $\bar{\epsilon}$ versus PEPS bond dimension $D$ for (a) the spin-1/2 frustrated $J_1$-$J_2$ Heisenberg model at $J_2/J_1=0.5$ on $16\times 16$ with PEPS $D=10$ energy as the reference ~\cite{liu2024tensor},  and (b) the 2D Hubbard model with onsite  repulsive interaction $U=8$ at $1/8$ hole doping with extrapolated DMRG energies as the reference~\cite{liu2024hubbard}. For both cases the computational cost scales as $O(D^6)$. The linear fits of $\log{\bar{\epsilon}} $ versus $\log {D}$ give  $\bar{\epsilon} \sim 1/D^{4}$ for the $J_1$-$J_2$ model (red dashed line) and  $\bar{\epsilon} \sim 1/D^{1.5}$ for the Hubbard model. Note that the cost is $O(D^6)$, thus the range of $D$ considered spans a wide range of cost. }
  \label{fig:tns}
\end{figure}

Focusing then on the strongly correlated limit, we now consider systems for which a single mean-field reference is certainly a poor qualitative description. In Fig.~\ref{fig:tns}, we show the error convergence of a tensor network ansatz (the projected entangled pair state, PEPS) for the 2D frustrated $J_1$-$J_2$ Heisenberg model, and the 4-leg and 8-leg doped 2D Hubbard models. The data on the $J_1$-$J_2$ model is analyzed here for the first time, as is that for the 2D Hubbard model for the 8-leg ladder. Additional data, for the 3D Heisenberg model, can be found in Ref.~\cite{expadvantage}.

We find that as a function of the flexibility of the ansatz, as expressed by the tensor bond dimension $D$, the relative error behaves like $\bar{\epsilon} \sim \mathrm{poly}(1/D)$. 
Then since the computational cost is $O(L\mathrm{poly}(D))$, more precisely, $O(LD^6)$ in the variational Monte Carlo PEPS formulation used here~\cite{liu2021accurate}, the overall cost of refinement is $O(L /\bar{\epsilon}^{1.5})$ in the $J_1$-$J_2$ Heisenberg model, and $O(L /\bar{\epsilon}^{4})$ in the 2D Hubbard model, both consistent with $O(\mathrm{poly}(L)\mathrm{poly}(1/\epsilon))$ refinement cost. However, as observed in the coupled cluster examples, there is a range of exponents observed, consistent with the general understanding that fermionic systems are harder to simulate than spin systems with tensor network methods.

The above constitutes a very limited set of examples. In particular, while the calculations on strongly correlated models with tensor networks, and those on molecular systems with coupled cluster methods, correspond to two limits of strong and weak correlation (and are also representative of material versus finite molecular systems) we do not have much data for strongly correlated ab initio problems, where one needs to converge the description of both strong and weak correlation. This reflects the gap in current methodologies discussed in Sec.~\ref{sec:survey}. 
Nonetheless, if we take the evidence in hand, and with a certain degree of optimism that this author possesses, 
we can argue that the empirical error convergence, when using the appropriate classical heuristics for the given problem to refine the energy, will be $O(\mathrm{poly}(L)\mathrm{poly}(1/\epsilon))$, although it is likely that in some problems there is a rather high power dependence on the inverse error.  


\subsection{Summary}
\label{sec:classicalsummary}

This section has attempted to formalize the intuition expressed in the earlier part of this essay into some statements about the empirical complexity of using quantum chemistry heuristics. We now briefly summarize the key points. 

Classical heuristics are methods that are used with $O(\mathrm{poly}(L))$ runtimes, essentially by definition. However, the errors from running the heuristics will need to better characterized for a full understanding of their effectiveness. 

It appears to be useful to separate the error into two categories: a reference error, and a refinement error. We conjecture that the reference error can be eliminated by simply constructing an appropriate reference, which is some classically succinct state. Often, constructing such an appropriate reference is easy (e.g. one can just take a mean-field reference), but there are also problems where finding the reference will involve a classical search over a set of possibilities, which may be exponentially large in some difficult problems. 
The search over reference states may be where the exponential complexity of quantum chemistry ultimately lies, but our conjecture suggests that it is a type of classical exponential hardness, rather than a quantum exponential hardness.  

Once a suitable reference is identified, limited numerical data suggests that the refinement can be done efficiently (although perhaps not practically) by current classical heuristics. 
The cost of classical approaches under these assumptions thus takes the conjectured form $R_\text{state prep} \times O(\mathrm{poly}({L})\mathrm{poly}(1/\epsilon))$. We  refer to this hence as the \emph{classical heuristic cost conjecture}.

\section{Quantum algorithms and quantum advantage in quantum chemistry}

So far, this essay has focused only on classical algorithms for quantum chemistry, as these are methods that can be applied today and form the basis for existing chemical intuition. However, as we look towards the future, we can seek to understand the potential role of quantum algorithms in the problems of simulating chemical matter. While we can only touch on the surface of this emerging field and limit ourselves  to the problem of electronic structure, we will attempt to present some sense of where one can look for quantum advantage assuming that quantum chemistry follows the intuitions conjectured above. We will assume throughout that the quantum hardware we are using
is perfect (i.e. fault tolerant or fully error corrected).

To formalize the notion of advantage, 
we define the computational task as that of finding the ground-state energy of a many-electron quantum chemical problem of size $L$ (for the kinds of chemical Hamiltonians we have been discussing) to some desired error $\epsilon$. Then,  quantum advantage refers to a favourable relationship between the classical and quantum costs for this task. Usually, the most desirable advantage and the one most discussed is an asymptotic exponential quantum advantage (EQA)~\cite{expadvantage}. 
One way to define advantage is as the ratio of costs, then EQA means that asymptotically the ratio  is $e^{O(L)}$. EQA is clearly achieved if the quantum algorithm has polynomial cost but the classical algorithm has exponential cost, but under this definition, if the classical and quantum algorithms both have exponential cost, we still obtain an exponential advantage if the exponent in the quantum case is smaller. A different way to assess advantage is to define the classical cost as a function of the quantum cost, i.e. $C_\text{classical} = f(C_\text{quantum})$ and ask if this is a polynomial or exponential function. Under this definition, if say $C_\text{classical} = e^{cL}$ and $C_\text{quantum} = e^{cL/2}$, one would say that there is quadratic speedup and no EQA. The latter definition is probably more common (but note that Ref.~\cite{expadvantage} uses the first definition). We also note that there are other kinds of advantage, such as polynomial quantum advantage or even constant factor advantage.

In principle, under common complexity assumptions, quantum computation provides a strict superset of classical computational power. Thus, assuming a quantum computer has the option to execute the classical algorithm at the same speed as the classical computer, there must always be an asymptotic theoretical advantage to using a quantum computer. 
However, there is more to establishing quantum advantage in the quantum chemistry setting than this theoretical statement because we are generally interested only up to finite sizes and finite errors, as discussed in Sec.~\ref{sec:asymptotic}. As a consequence, the detailed form of speedup, including the prefactors, is important. While asymptotic analysis to date has usually ignored these details,  in part because some of the subleading costs result from the implementations of the algorithms on hardware that has yet to exist, it is clear that non-asymptotic analysis must play an important role in demonstrating quantum advantage in the quantum chemistry.

There are currently two main kinds of algorithms that have been proposed for simulating the quantum chemistry problem (i) variational (or hybrid) algorithms and (ii) algorithms based on phase estimation or quantum linear algebra. Although the first class of algorithms is often discussed in the context of noisy, non-error-corrected quantum computers, we assume below they are executed on perfect quantum hardware. 


\subsection{How to search for quantum advantage in chemical matter}

\label{sec:eqa}

As summarized in Sec.~\ref{sec:classicalsummary}, we argue that empirical observations suggest that classical quantum chemistry methods satisfy the classical heuristic cost conjecture, namely, they can be applied with an empirical complexity 
$R_\text{state prep} \times C_\text{refinement}$, where the first takes the form of a search, and the latter is of the form $O(\mathrm{poly}(L)\mathrm{poly}(1/\epsilon))$.  For classes of problems where $R_\text{state prep} = O(1)$ (e.g. the organic systems considered in Ref.~\cite{expadvantage}) the classical heuristic cost conjecture leaves no apparent room for an asymptotic exponential quantum advantage. However, whereas the asymptotic $O(\mathrm{poly}(L))$ component of the classical cost is strongly supported by the empirical observation of the local behaviour of matter, from the perspective of searching for quantum advantage, the polynomial inverse error dependence of classical heuristics has much weaker support as a conjecture and is thus a good place to look.

Indeed, it is clear that not every plausible classical heuristic has polynomial inverse error dependence, even under strong assumptions about the behaviour of chemical matter. For example, in a system exhibiting strict locality, one might choose to simulate the system as a set of non-overlapping fragments and add the fragment energies together; and one could, for example, use a full configuration interaction solver to simulate the ground state of each fragment. For any fixed fragment size $M$, the scaling of the method is then $O(\mathrm{poly}(L))$ in system size (the polynomial dependence above linear arising from the long-range Coulomb interaction) but each fragment can be expected to have an energy error proportional to its surface to volume ratio, and since the cost to compute a fragment's energy when using a full configuration interation solver is $e^{O(M)}$, the relative error scaling of the method becomes $e^{O(1/\bar{\epsilon})}$, translating to an overall total cost of ${e^{O(L)} e^{O(1/\epsilon)}}$ when considering the absolute error. This then provides room for a quantum algorithm to achieve EQA. Obtaining  $O(\mathrm{poly}(1/\epsilon))$ error clearly requires a careful heuristic approach, and we cannot prove that such a heuristic exists for all chemical matter (although we also have yet to identify concrete counter-examples of a chemically relevant nature~\cite{expadvantage}). 

Even if the classical heuristic cost conjecture holds, i.e. we can always find methods with $O(\mathrm{poly}(1/\epsilon))$ refinement error, the polynomial power may also be very large. This is seen in some of the numerical examples in Sec.~\ref{sec:classicalrefinement} and presents a situation where a quantum algorithm can potentially achieve a large polynomial speedup. Given that we should always consider there to be a finite size cutoff $L_c$ in chemistry anyways, a large polynomial speedup up to sizes of $O(L_c)$ might be viewed as a more appropriate target than asymptotic exponential quantum advantage. 

There is also the class of problems where $R_\text{state prep} \sim e^{O(L)}$. For these problems, there is again the possibility that the subsequent classical refinement is inefficient, leading to a similar mechanism for quantum speedup as described above. But assuming the classical energy can be refined efficiently, then another possibility is that there is a quantum speedup that arises from eliminating the repeated classical state preparation. However, in the author's view, problems of this kind arise mainly when there are competing references (competing phases in a material) and such problems have characteristics of spin glasses, where it is already known that quantum algorithms for ground states must themselves have a similar exponential cost, essentially due to a similar state preparation problem. 
Spin glasses offer no room for exponential speedup arising from the state preparation, although EQA (in terms of the first definition) is still possible in terms of a ratio of exponential costs. While one might search for more exotic matter where the quantum state preparation problem is exponentially easier than the classical one (and there are certainly artificial Hamiltonians where this can be made to be the case~\cite{chen2024local}) we do not know of a chemically relevant example, and arguably, finding such a fine-tuned case has limited impact in the discipline of quantum chemistry.

\subsection{Conjectured sources of quantum advantage from common to uncommon}

From the above analysis, we can  conjecture what the different manifestations of quantum advantage in quantum chemistry simulations of chemical matter look like. The most common case will be systems where the classical heuristic cost conjecture of this essay holds; then by understanding the relative error convergence of classical versus quantum algorithms across chemical matter, we can find a subset where a polynomial advantage can be achieved. The second case requires a failure of the refinement part of the cost conjecture, namely we must identify problems where there is a classically succinct reference, but the classical refinement cannot be efficiently performed, thus giving the possibility for an exponential quantum advantage. The final case is where the speedup arises primarily from state preparation. A large advantage in this case would  require finding problems where a classically succinct reference does not exist, or where quantum state preparation is much easier than classical state preparation. These different sources of speedup are summarized in Table~\ref{tab:speedups}.

\begin{table}\centering
\begin{tabular}{c|c|c|c}
\hline\hline
Classical & Quantum speedup & Advantage & Likelihood \\
\hline
 $R_\text{state prep} \times O(\mathrm{poly}(L)\mathrm{poly}(1/\epsilon))$ &  refinement & poly. & common \\
 $R_\text{state prep} \times O(\mathrm{poly}(L)\exp{(O(1/\bar{\epsilon}))}$ &  refinement & exp. & possible \\
 $e^{O(L)} \times C_\text{refinement}$ & state prep. & poly. & common \\
 $e^{O(L)} \times C_\text{refinement}$ &  state prep. & exp. & unlikely \\
     \hline\hline
\end{tabular}
\caption{Some conjectured manifestations of quantum speedup in quantum chemistry simulations and their proposed likelihood across problems of chemical relevance. Advantage here refers to the functional form of $C_\text{classical} = f(C_\text{quantum})$. $\bar{\epsilon}=\epsilon/L$.
\label{tab:speedups}}
\end{table}

\subsection{Advantage in hybrid and variational quantum algorithms}

Hybrid algorithms are quantum algorithms which contain an important element of classical computation in conjunction with quantum operations: the classical and quantum computing hardware are expected to exchange a (small amount) of classical data in a self-consistent loop~\cite{peruzzo2014variational,motta2020determining,bauer2020quantum}. The largest family of such algorithms are the variational algorithms~\cite{peruzzo2014variational}, which perform the minimization
\begin{align}
  E = \min_\theta \langle 0|U^\dag(\theta) H U(\theta)|0\rangle \label{eq:vqe}
\end{align}
where $U(\theta)$ is a parametrized quantum circuit, $\theta$ are the variational parameters, and $|0\rangle$ is some simple initial state on the quantum device. The energy is measured (with statistical shot noise) on the quantum device by expressing $H$ as a sum of terms (i.e. individual second quantized operators, such as $a^\dag_i a^\dag_j a_k a_l$, which, under some particular fermion encoding, is represented as a Pauli string) and the total energy is minimized.

Variational algorithms, such as the variational quantum eigensolver~\cite{peruzzo2014variational}, have elicited an enormous amount of attention and are extremely popular. To some extent, this can be attributed to the fact that are easy to understand: any circuit parametrization yields a viable ansatz which may even be tested on one of today's existing noisy quantum devices. On the other hand, they contain a non-linear optimization and involve very large prefactors due to the number of measurements required to reduce the statistical error (for some examples of numbers, see~\cite{peng2023fermionic}). Even disregarding these constant prefactors it is currently difficult to find quantitative evidence of quantum advantage with variational algorithms in problems of chemical matter. In principle, a circuit ansatz must  be able to represent states that cannot   be efficiently represented on a classical device. However, this by itself does not mean there is a quantum advantage, because one needs to (i) first find the circuit representation that yields such a state through the optimization in Eq.~\ref{eq:vqe} and (ii) ensure that there is not some other classical representation of the wavefunction that yields a similar quality of the energy or other observable, with a lower cost.

The first issue has been much discussed in the quantum literature because, in general, optimizing unitary circuits is difficult, due to the ``barren plateau'' problem, where a gradient algorithm gets stuck in some high-energy local minimum~\cite{larocca2024review}. Indeed, this has led some to question the value of variational quantum algorithms~\cite{cerezo2023does}.
Here we might appeal to the classical intuitions established above to ameliorate this problem, namely we can initialize with the same classically succinct reference states used in classical quantum chemistry: this might be viewed as the state preparation step of a variational quantum algorithm. Whether this solves or ameliorates the barren plateau problem remains to be established. Under the assumption that it does, there is still a possibility for polynomial quantum advantage with respect to the search over classical references (although the size of this advantage may be small). Given the above strategy, the cost of a variational algorithm may be analyzed as
\begin{align}
    C = R_\text{init} \times C_\text{refinement}
\end{align}
where $R_\text{init}$ is the number of different initial guesses (here assumed to be classically succinct states) that must be tried.



We can then assess the possibility of quantum advantage using a variational quantum circuit for energy refinement. One study which attempts to do this is Ref.~\cite{variationaltns}, which compares variational quantum algorithms with different circuit architectures to classical MPS calculations on 1D (and pseudo-1D) model Hamiltonians.
Because DMRG and MPS are recognized as some of the best classical algorithms in the 1D context, we can argue that this is a reasonable classical comparison.  Ref.~\cite{variationaltns} examined the expressivity (the number of variational parameters) and the theoretical cost needed to reach a desired accuracy in the ground-state energy. Extrapolating to high accuracy, and neglecting the difficulty of optimization, it showed that for certain kinds of circuit ansatz, there is a small polynomial advantage in the best circuit ansatz considered relative to a standard classical MPS. {For example, for the standard MPS, the theoretical cost in the Heisenberg model (for a rectangular strip, fixed system size) to obtain a relative precision of $\bar{\epsilon}$ was estimated to be roughly $C \sim O(1/{\bar{\epsilon}^{3.1})}$, while with the best circuit structure, it was found that $C \sim O(1/\bar{\epsilon}^{2.9})$.} 

The above study provides partial evidence of a very modest polynomial quantum advantage in variational quantum algorithms coming from the error refinement task. 
But it should also be noted that the advantage is achieved in part because the chosen quantum circuit is a more structured and restricted ansatz than the classical MPS to which it is compared and it remains to be seen whether analogous simplifications used in the classical setting eliminate this advantage.
Further studies of this kind, which compare the asymptotic expressiveness and computational costs of different variational circuits against competing classical methods, are clearly needed.

\subsection{Advantage in phase estimation and quantum linear algebra based algorithms}

The second class of algorithms are ones that use some form of quantum linear algebra or quantum phase estimation to obtain the ground-state. We will focus on quantum phase estimation~\cite{nielsen2010quantum} since the complexity is representative (up to polynomial factors~\cite{lin2020near}) of these approaches. The total cost of quantum phase estimation  for the ground state is of the form
\begin{align}
  C = R_\text{state prep} \times C_\text{phase estimation}
\end{align}
where phase estimation (approximately) provides a projective measurement onto the ground-state of $H$ with a success probability depending on the quality of the initial state prepared. The cost can thus be written as
\begin{align}
  C \sim \mathrm{poly}(1/S) \times \mathrm{poly}(L) \times \mathrm{poly}(1/\epsilon) 
\end{align}
where $S = \langle \Phi|\Psi_0\rangle$, with $|\Phi\rangle$ being the initial state prepared.

Much has been made of the complexity of ground-state preparation for phase estimation because for an arbitrary Hamiltonian, it is known that preparing a state with good overlap with the desired ground-state is generally hard even for a quantum computer (i.e. \textsc{QMA} hard). But, assuming the intuitions we have argued above, this is a red herring for chemical matter: no problem of real-world chemical relevance appears to have QMA hard state preparation complexity.

As an aside, we note that preparing a state for phase estimation is both a harder and (potentially) an easier task than preparing a standard classical reference state. It is a harder task because one wants to prepare a state with asymptotically large overlap with the ground-state (i.e. at worst $O(\mathrm{poly}(1/L))$, while the usual standard classical reference states (such as a mean-field state, a coupled cluster wavefunction with fixed truncation order, or a tensor network state of some fixed bond dimension) actually have exponentially small overlap with the true state as $L \to \infty$, giving exponential cost in the phase estimation algorithm from the $\mathrm{poly}(1/S)$ factor. Thus one needs to prepare an initial state that is asymptotically better than standard classical reference states. 

On the other hand it is potentially an easier task, because the task of preparing a state of good overlap could, in principle, be easier than finding a reference from which a classical heuristic can easily refine an energy. To see an example of this, we can consider the difficulty of preparing a classical reference state such that perturbation theory converges, compared to that of performing adiabatic state preparation. In the perturbation theory case, we consider the Hamiltonian $H(\lambda) = H_0 + \lambda H_1$ (with the desired Hamiltonian corresponding to $\lambda=1$) and the energy $E(\lambda)$. Then for perturbation theory to converge, $E(\lambda)$
must be analytic in the complex circle for $|\lambda| < 1$, which requires the gap not to close in the complex circle. On the other hand, in the case of adiabatic state preparation, we consider an adiabatic path such as $H(\lambda) = (1-\lambda) H_0 + \lambda H_1$, and we only require that the gap does not close along the path, a presumably weaker requirement. 

However, in Ref.~\cite{expadvantage}, we suggest that at least in chemical matter, the difficulty of state preparation for classical heuristics and for phase estimation are closely related. For example, systems in which a classical search over reference states needs to be performed to find a reference for  classical heuristics (such as the Fe-S clusters in nitrogenase), appear to engender a similar level of complexity in quantum state preparation methods for phase estimation, such as adiabatic state preparation. On the other hand once a suitable classical reference for classical heuristics is known (which can be expected to have exponentially small overlap for large systems) the subsequent application of classical refinement not only rapidly improves the energy, but also rapidly improves the overlap, yielding a state that may then be prepared on a quantum computer for phase estimation.

Moving on from the state preparation aspect, we now consider the energy refinement which is performed by the phase estimation measurement. Phase estimation gives what is termed the Heisenberg limited error cost $O(1/\epsilon)$ - and this is generically the best error scaling one can get for a quantum algorithm. Under the classical heuristic cost conjecture, we assume a classical error scaling of $O(\mathrm{poly}(1/\epsilon))$, but it is possible for the polynomial exponent to be both $<1$ and thus better than what is seen in phase estimation (e.g. as observed in the 1D Heisenberg model in Ref.~\cite{haghshenas2021numerical}) and also much greater than 1, as seen in some of the examples in Sec.~\ref{sec:classicalrefinement}. 

Overall, however, the well-understood and provable error scaling of quantum phase estimation provides a potential source of  polynomial quantum advantage in a wide range of chemical matter even under the assumptions of the classical heuristic cost conjecture (and of course potentially even more advantage if the conjecture breaks down). Given the diversity of the polynomial inverse error dependence seen in classical simulations, establishing this advantage concretely will require studying a wide range of examples.  Because of the non-asymptotic nature of the actual chemical problems of chemical matter, the quantum algorithms must also be fully characterized with respect to their non-asymptotic costs.

\subsection{Summary}

In the above, we have considered how to search for quantum advantage under the constraints of chemical matter. We can summarize these constraints in the form of the classical heuristic cost conjecture introduced in Sec.~\ref{sec:classicalsummary}. Then searching for quantum advantage requires understanding the types of advantage that are available when (and if) the conjecture holds, as well as the ways in which the conjecture can fail, and the subsequent advantages that then appear. With respect to the failure of the classical heuristic cost conjecture,  we argue that it is useful to distinguish between two modes; failure of the classical state preparation and failure of the classical refinement. For the widely applicable quantum advantage in chemically relevant systems, we suggest that focusing on quantum advantage in refinement will be key.

We apply this reasoning to
two families of quantum algorithms, variational quantum algorithms and quantum phase estimation (and related quantum algorithms). Both can be formulated in terms of a state preparation piece and a refinement component, thus creating a parallel functional form of the cost between classical heuristics, variational quantum algorithms, and quantum phase estimation.
While the error convergence of variational quantum algorithms must be better established to understand the advantage that can be achieved, algorithms such as phase estimation, through their rigorous Heisenberg limited error scaling, present a strong case for looking for quantum advantage in refinement. Certainly we can expect to see polynomial speedups in some problems, and searching for a breakdown of classical refinement and the classical heuristic cost conjecture is an avenue to exponential speedup.
However, understanding the specific type of advantage will require analysis beyond asymptotics.

\section{Conclusions}

In this essay we have attempted to provide an overview of quantum chemistry, surveying the problems, methods, intuitions, and complexity of both classical and quantum methods. For the quantum chemistry reader, we hope this essay has illustrated some new ways to think about the computational complexity of chemical problems that may help to solidify existing intuitions and lead to new ones.  For the quantum information theory reader, we hope this work has provided a guide to the diversity of chemical problems and the rich lore of chemical intuition that has laid the  foundations of the quantum chemistry field. 

Ultimately, chemistry is an experimental science, and the domain of its problems and the associated complexities has been defined not by mathematics, but by the examples in Nature and our experimental control. In formulating the conjectures of this essay, we have tried to capture some essence of  the behaviour of chemical matter that is amenable to mathematical analysis. We look forward to new developments in these directions.

\section*{Acknowledgements}
This was a long article to write and the author  is indebted to the generous time of many people who  read parts of the manuscript and offered their opinions. We would like to acknowledge useful discussions with Alex Dalzell, Sam McArdle, Ryan Babbush, Steve White, Ulf Ryde, Sandeep Sharma, Ed Valeev, Tim Berkelbach, Lin Lin, Jiaqing Jiang, Jonas Peters, Doug Rees, and Karthish Manthiraram. In addition I am grateful to the contributions of the members of my group, in particular, Chris Chen, Huanchen Zhai, Yuhang Ai, Chenghan Li, Gunhee Park, Verena Neufeld, Yu Tong, and Wenyuan Liu, who proofread, prepared figures, fact checked, and provided other useful suggestions.

The author acknowledges support from the U.S. Department of Energy, Office of Science, National Quantum Information Science Research
Centers, Quantum Systems Accelerator.




\footnotesize{
\bibliography{rsc} 

\providecommand*{\mcitethebibliography}{\thebibliography}
\csname @ifundefined\endcsname{endmcitethebibliography}
{\let\endmcitethebibliography\endthebibliography}{}
\begin{mcitethebibliography}{73}
\providecommand*{\natexlab}[1]{#1}
\providecommand*{\mciteSetBstSublistMode}[1]{}
\providecommand*{\mciteSetBstMaxWidthForm}[2]{}
\providecommand*{\mciteBstWouldAddEndPuncttrue}
  {\def\EndOfBibitem{\unskip.}}
\providecommand*{\mciteBstWouldAddEndPunctfalse}
  {\let\EndOfBibitem\relax}
\providecommand*{\mciteSetBstMidEndSepPunct}[3]{}
\providecommand*{\mciteSetBstSublistLabelBeginEnd}[3]{}
\providecommand*{\EndOfBibitem}{}
\mciteSetBstSublistMode{f}
\mciteSetBstMaxWidthForm{subitem}
{(\emph{\alph{mcitesubitemcount}})}
\mciteSetBstSublistLabelBeginEnd{\mcitemaxwidthsubitemform\space}
{\relax}{\relax}

\bibitem[Arora and Barak(2009)]{arora2009computational}
S.~Arora and B.~Barak, \emph{Computational complexity: a modern approach}, Cambridge University Press, 2009\relax
\mciteBstWouldAddEndPuncttrue
\mciteSetBstMidEndSepPunct{\mcitedefaultmidpunct}
{\mcitedefaultendpunct}{\mcitedefaultseppunct}\relax
\EndOfBibitem
\bibitem[Dirac(1929)]{dirac1929quantum}
P.~A.~M. Dirac, \emph{Proceedings of the Royal Society of London. Series A, Containing Papers of a Mathematical and Physical Character}, 1929, \textbf{123}, 714--733\relax
\mciteBstWouldAddEndPuncttrue
\mciteSetBstMidEndSepPunct{\mcitedefaultmidpunct}
{\mcitedefaultendpunct}{\mcitedefaultseppunct}\relax
\EndOfBibitem
\bibitem[Ryde and Soderhjelm(2016)]{ryde2016ligand}
U.~Ryde and P.~Soderhjelm, \emph{Chemical Reviews}, 2016, \textbf{116}, 5520--5566\relax
\mciteBstWouldAddEndPuncttrue
\mciteSetBstMidEndSepPunct{\mcitedefaultmidpunct}
{\mcitedefaultendpunct}{\mcitedefaultseppunct}\relax
\EndOfBibitem
\bibitem[Gilli \emph{et~al.}(1994)Gilli, Ferretti, Gilli, and Borea]{gilli1994enthalpy}
P.~Gilli, V.~Ferretti, G.~Gilli and P.~A. Borea, \emph{The Journal of Physical Chemistry}, 1994, \textbf{98}, 1515--1518\relax
\mciteBstWouldAddEndPuncttrue
\mciteSetBstMidEndSepPunct{\mcitedefaultmidpunct}
{\mcitedefaultendpunct}{\mcitedefaultseppunct}\relax
\EndOfBibitem
\bibitem[Rouwenhorst \emph{et~al.}(2021)Rouwenhorst, Krzywda, Benes, Mul, and Lefferts]{ROUWENHORST202141}
K.~Rouwenhorst, P.~Krzywda, N.~Benes, G.~Mul and L.~Lefferts, \emph{Techno-Economic Challenges of Green Ammonia as an Energy Vector}, Academic Press, 2021, pp. 41--83\relax
\mciteBstWouldAddEndPuncttrue
\mciteSetBstMidEndSepPunct{\mcitedefaultmidpunct}
{\mcitedefaultendpunct}{\mcitedefaultseppunct}\relax
\EndOfBibitem
\bibitem[Milo and Phillips(2015)]{milo2015cell}
R.~Milo and R.~Phillips, \emph{Cell biology by the numbers}, Garland Science, 2015\relax
\mciteBstWouldAddEndPuncttrue
\mciteSetBstMidEndSepPunct{\mcitedefaultmidpunct}
{\mcitedefaultendpunct}{\mcitedefaultseppunct}\relax
\EndOfBibitem
\bibitem[Sharma \emph{et~al.}(2014)Sharma, Sivalingam, Neese, and Chan]{sharma2014low}
S.~Sharma, K.~Sivalingam, F.~Neese and G.~K.-L. Chan, \emph{Nature chemistry}, 2014, \textbf{6}, 927--933\relax
\mciteBstWouldAddEndPuncttrue
\mciteSetBstMidEndSepPunct{\mcitedefaultmidpunct}
{\mcitedefaultendpunct}{\mcitedefaultseppunct}\relax
\EndOfBibitem
\bibitem[Li \emph{et~al.}(2019)Li, Guo, Sun, and Chan]{li2019electronic}
Z.~Li, S.~Guo, Q.~Sun and G.~K.-L. Chan, \emph{Nature chemistry}, 2019, \textbf{11}, 1026--1033\relax
\mciteBstWouldAddEndPuncttrue
\mciteSetBstMidEndSepPunct{\mcitedefaultmidpunct}
{\mcitedefaultendpunct}{\mcitedefaultseppunct}\relax
\EndOfBibitem
\bibitem[Bjornsson \emph{et~al.}(2015)Bjornsson, Neese, Schrock, Einsle, and DeBeer]{bjornsson2015discovery}
R.~Bjornsson, F.~Neese, R.~R. Schrock, O.~Einsle and S.~DeBeer, \emph{JBIC Journal of Biological Inorganic Chemistry}, 2015, \textbf{20}, 447--460\relax
\mciteBstWouldAddEndPuncttrue
\mciteSetBstMidEndSepPunct{\mcitedefaultmidpunct}
{\mcitedefaultendpunct}{\mcitedefaultseppunct}\relax
\EndOfBibitem
\bibitem[Jiang \emph{et~al.}(2023)Jiang, Lundgren, and Ryde]{jiang2023protonation}
H.~Jiang, K.~J. Lundgren and U.~Ryde, \emph{Inorganic Chemistry}, 2023, \textbf{62}, 19433--19445\relax
\mciteBstWouldAddEndPuncttrue
\mciteSetBstMidEndSepPunct{\mcitedefaultmidpunct}
{\mcitedefaultendpunct}{\mcitedefaultseppunct}\relax
\EndOfBibitem
\bibitem[Zheng \emph{et~al.}(2017)Zheng, Chung, Corboz, Ehlers, Qin, Noack, Shi, White, Zhang, and Chan]{zheng2017stripe}
B.-X. Zheng, C.-M. Chung, P.~Corboz, G.~Ehlers, M.-P. Qin, R.~M. Noack, H.~Shi, S.~R. White, S.~Zhang and G.~K.-L. Chan, \emph{Science}, 2017, \textbf{358}, 1155--1160\relax
\mciteBstWouldAddEndPuncttrue
\mciteSetBstMidEndSepPunct{\mcitedefaultmidpunct}
{\mcitedefaultendpunct}{\mcitedefaultseppunct}\relax
\EndOfBibitem
\bibitem[Arovas \emph{et~al.}(2022)Arovas, Berg, Kivelson, and Raghu]{arovas2022hubbard}
D.~P. Arovas, E.~Berg, S.~A. Kivelson and S.~Raghu, \emph{Annual review of condensed matter physics}, 2022, \textbf{13}, 239--274\relax
\mciteBstWouldAddEndPuncttrue
\mciteSetBstMidEndSepPunct{\mcitedefaultmidpunct}
{\mcitedefaultendpunct}{\mcitedefaultseppunct}\relax
\EndOfBibitem
\bibitem[Xu \emph{et~al.}(2024)Xu, Chung, Qin, Schollw{\"o}ck, White, and Zhang]{xu2024coexistence}
H.~Xu, C.-M. Chung, M.~Qin, U.~Schollw{\"o}ck, S.~R. White and S.~Zhang, \emph{Science}, 2024, \textbf{384}, eadh7691\relax
\mciteBstWouldAddEndPuncttrue
\mciteSetBstMidEndSepPunct{\mcitedefaultmidpunct}
{\mcitedefaultendpunct}{\mcitedefaultseppunct}\relax
\EndOfBibitem
\bibitem[Weber \emph{et~al.}(2010)Weber, Haule, and Kotliar]{weber2010strength}
C.~Weber, K.~Haule and G.~Kotliar, \emph{Nature Physics}, 2010, \textbf{6}, 574--578\relax
\mciteBstWouldAddEndPuncttrue
\mciteSetBstMidEndSepPunct{\mcitedefaultmidpunct}
{\mcitedefaultendpunct}{\mcitedefaultseppunct}\relax
\EndOfBibitem
\bibitem[Schmid \emph{et~al.}(2023)Schmid, Mor{\'e}e, Kaneko, Yamaji, and Imada]{schmid2023superconductivity}
M.~T. Schmid, J.-B. Mor{\'e}e, R.~Kaneko, Y.~Yamaji and M.~Imada, \emph{Physical Review X}, 2023, \textbf{13}, 041036\relax
\mciteBstWouldAddEndPuncttrue
\mciteSetBstMidEndSepPunct{\mcitedefaultmidpunct}
{\mcitedefaultendpunct}{\mcitedefaultseppunct}\relax
\EndOfBibitem
\bibitem[Cui \emph{et~al.}(2022)Cui, Zhai, Zhang, and Chan]{cui2022systematic}
Z.-H. Cui, H.~Zhai, X.~Zhang and G.~K.-L. Chan, \emph{Science}, 2022, \textbf{377}, 1192--1198\relax
\mciteBstWouldAddEndPuncttrue
\mciteSetBstMidEndSepPunct{\mcitedefaultmidpunct}
{\mcitedefaultendpunct}{\mcitedefaultseppunct}\relax
\EndOfBibitem
\bibitem[Cui \emph{et~al.}(2023)Cui, Yang, T{\"o}lle, Ye, Zhai, Kim, Zhang, Lin, Berkelbach, and Chan]{cui2023ab}
Z.-H. Cui, J.~Yang, J.~T{\"o}lle, H.-Z. Ye, H.~Zhai, R.~Kim, X.~Zhang, L.~Lin, T.~C. Berkelbach and G.~K. Chan, \emph{arXiv preprint arXiv:2306.16561}, 2023\relax
\mciteBstWouldAddEndPuncttrue
\mciteSetBstMidEndSepPunct{\mcitedefaultmidpunct}
{\mcitedefaultendpunct}{\mcitedefaultseppunct}\relax
\EndOfBibitem
\bibitem[Babbush \emph{et~al.}(2018)Babbush, Wiebe, McClean, McClain, Neven, and Chan]{babbush2018low}
R.~Babbush, N.~Wiebe, J.~McClean, J.~McClain, H.~Neven and G.~K.-L. Chan, \emph{Physical Review X}, 2018, \textbf{8}, 011044\relax
\mciteBstWouldAddEndPuncttrue
\mciteSetBstMidEndSepPunct{\mcitedefaultmidpunct}
{\mcitedefaultendpunct}{\mcitedefaultseppunct}\relax
\EndOfBibitem
\bibitem[Kong \emph{et~al.}(2012)Kong, Bischoff, and Valeev]{kong2012explicitly}
L.~Kong, F.~A. Bischoff and E.~F. Valeev, \emph{Chemical reviews}, 2012, \textbf{112}, 75--107\relax
\mciteBstWouldAddEndPuncttrue
\mciteSetBstMidEndSepPunct{\mcitedefaultmidpunct}
{\mcitedefaultendpunct}{\mcitedefaultseppunct}\relax
\EndOfBibitem
\bibitem[Mardirossian and Head-Gordon(2017)]{mardirossian2017thirty}
N.~Mardirossian and M.~Head-Gordon, \emph{Molecular physics}, 2017, \textbf{115}, 2315--2372\relax
\mciteBstWouldAddEndPuncttrue
\mciteSetBstMidEndSepPunct{\mcitedefaultmidpunct}
{\mcitedefaultendpunct}{\mcitedefaultseppunct}\relax
\EndOfBibitem
\bibitem[Jensen(2007)]{jensen2007}
F.~Jensen, \emph{Introduction to Computational Chemistry}, John Wiley \& Sons, 2nd edn, 2007\relax
\mciteBstWouldAddEndPuncttrue
\mciteSetBstMidEndSepPunct{\mcitedefaultmidpunct}
{\mcitedefaultendpunct}{\mcitedefaultseppunct}\relax
\EndOfBibitem
\bibitem[Ismail-Beigi and Arias(1999)]{ismail1999locality}
S.~Ismail-Beigi and T.~A. Arias, \emph{Physical review letters}, 1999, \textbf{82}, 2127\relax
\mciteBstWouldAddEndPuncttrue
\mciteSetBstMidEndSepPunct{\mcitedefaultmidpunct}
{\mcitedefaultendpunct}{\mcitedefaultseppunct}\relax
\EndOfBibitem
\bibitem[Sandvik(1997)]{sandvik1997finite}
A.~W. Sandvik, \emph{Physical Review B}, 1997, \textbf{56}, 11678\relax
\mciteBstWouldAddEndPuncttrue
\mciteSetBstMidEndSepPunct{\mcitedefaultmidpunct}
{\mcitedefaultendpunct}{\mcitedefaultseppunct}\relax
\EndOfBibitem
\bibitem[Eisert \emph{et~al.}(2010)Eisert, Cramer, and Plenio]{eisert2010colloquium}
J.~Eisert, M.~Cramer and M.~B. Plenio, \emph{Reviews of modern physics}, 2010, \textbf{82}, 277--306\relax
\mciteBstWouldAddEndPuncttrue
\mciteSetBstMidEndSepPunct{\mcitedefaultmidpunct}
{\mcitedefaultendpunct}{\mcitedefaultseppunct}\relax
\EndOfBibitem
\bibitem[Brandao and Cramer(2015)]{brandao2015entanglement}
F.~G. Brandao and M.~Cramer, \emph{Physical Review B}, 2015, \textbf{92}, 115134\relax
\mciteBstWouldAddEndPuncttrue
\mciteSetBstMidEndSepPunct{\mcitedefaultmidpunct}
{\mcitedefaultendpunct}{\mcitedefaultseppunct}\relax
\EndOfBibitem
\bibitem[Helgaker \emph{et~al.}(2013)Helgaker, Jorgensen, and Olsen]{helgaker2013molecular}
T.~Helgaker, P.~Jorgensen and J.~Olsen, \emph{Molecular electronic-structure theory}, John Wiley \& Sons, 2013\relax
\mciteBstWouldAddEndPuncttrue
\mciteSetBstMidEndSepPunct{\mcitedefaultmidpunct}
{\mcitedefaultendpunct}{\mcitedefaultseppunct}\relax
\EndOfBibitem
\bibitem[Sharma \emph{et~al.}(2017)Sharma, Holmes, Jeanmairet, Alavi, and Umrigar]{sharma2017semistochastic}
S.~Sharma, A.~A. Holmes, G.~Jeanmairet, A.~Alavi and C.~J. Umrigar, \emph{Journal of chemical theory and computation}, 2017, \textbf{13}, 1595--1604\relax
\mciteBstWouldAddEndPuncttrue
\mciteSetBstMidEndSepPunct{\mcitedefaultmidpunct}
{\mcitedefaultendpunct}{\mcitedefaultseppunct}\relax
\EndOfBibitem
\bibitem[Loos \emph{et~al.}(2020)Loos, Damour, and Scemama]{loos2020performance}
P.-F. Loos, Y.~Damour and A.~Scemama, \emph{The Journal of Chemical Physics}, 2020, \textbf{153}, 176101\relax
\mciteBstWouldAddEndPuncttrue
\mciteSetBstMidEndSepPunct{\mcitedefaultmidpunct}
{\mcitedefaultendpunct}{\mcitedefaultseppunct}\relax
\EndOfBibitem
\bibitem[Cremer(2011)]{cremer2011moller}
D.~Cremer, \emph{Wiley Interdisciplinary Reviews: Computational Molecular Science}, 2011, \textbf{1}, 509--530\relax
\mciteBstWouldAddEndPuncttrue
\mciteSetBstMidEndSepPunct{\mcitedefaultmidpunct}
{\mcitedefaultendpunct}{\mcitedefaultseppunct}\relax
\EndOfBibitem
\bibitem[Olsen \emph{et~al.}(2000)Olsen, J{\o}rgensen, Helgaker, and Christiansen]{olsen2000divergence}
J.~Olsen, P.~J{\o}rgensen, T.~Helgaker and O.~Christiansen, \emph{The Journal of Chemical Physics}, 2000, \textbf{112}, 9736--9748\relax
\mciteBstWouldAddEndPuncttrue
\mciteSetBstMidEndSepPunct{\mcitedefaultmidpunct}
{\mcitedefaultendpunct}{\mcitedefaultseppunct}\relax
\EndOfBibitem
\bibitem[Martin \emph{et~al.}(2016)Martin, Reining, and Ceperley]{martin2016interacting}
R.~M. Martin, L.~Reining and D.~M. Ceperley, \emph{Interacting electrons}, Cambridge University Press, 2016\relax
\mciteBstWouldAddEndPuncttrue
\mciteSetBstMidEndSepPunct{\mcitedefaultmidpunct}
{\mcitedefaultendpunct}{\mcitedefaultseppunct}\relax
\EndOfBibitem
\bibitem[Shavitt and Bartlett(2009)]{shavitt2009many}
I.~Shavitt and R.~J. Bartlett, \emph{Many-body methods in chemistry and physics: MBPT and coupled-cluster theory}, Cambridge university press, 2009\relax
\mciteBstWouldAddEndPuncttrue
\mciteSetBstMidEndSepPunct{\mcitedefaultmidpunct}
{\mcitedefaultendpunct}{\mcitedefaultseppunct}\relax
\EndOfBibitem
\bibitem[Paldus \emph{et~al.}(1984)Paldus, Takahashi, and Cho]{paldus1984coupled}
J.~Paldus, M.~Takahashi and R.~Cho, \emph{Physical Review B}, 1984, \textbf{30}, 4267\relax
\mciteBstWouldAddEndPuncttrue
\mciteSetBstMidEndSepPunct{\mcitedefaultmidpunct}
{\mcitedefaultendpunct}{\mcitedefaultseppunct}\relax
\EndOfBibitem
\bibitem[Verstraete \emph{et~al.}(2008)Verstraete, Murg, and Cirac]{verstraete2008matrix}
F.~Verstraete, V.~Murg and J.~I. Cirac, \emph{Advances in physics}, 2008, \textbf{57}, 143--224\relax
\mciteBstWouldAddEndPuncttrue
\mciteSetBstMidEndSepPunct{\mcitedefaultmidpunct}
{\mcitedefaultendpunct}{\mcitedefaultseppunct}\relax
\EndOfBibitem
\bibitem[Chan \emph{et~al.}(2004)Chan, K{\'a}llay, and Gauss]{chan2004state}
G.~K.-L. Chan, M.~K{\'a}llay and J.~Gauss, \emph{The Journal of Chemical Physics}, 2004, \textbf{121}, 6110--6116\relax
\mciteBstWouldAddEndPuncttrue
\mciteSetBstMidEndSepPunct{\mcitedefaultmidpunct}
{\mcitedefaultendpunct}{\mcitedefaultseppunct}\relax
\EndOfBibitem
\bibitem[Hastings(2007)]{hastings2007area}
M.~B. Hastings, \emph{Journal of statistical mechanics: theory and experiment}, 2007, \textbf{2007}, P08024\relax
\mciteBstWouldAddEndPuncttrue
\mciteSetBstMidEndSepPunct{\mcitedefaultmidpunct}
{\mcitedefaultendpunct}{\mcitedefaultseppunct}\relax
\EndOfBibitem
\bibitem[White(1993)]{white1993density}
S.~R. White, \emph{Physical review b}, 1993, \textbf{48}, 10345\relax
\mciteBstWouldAddEndPuncttrue
\mciteSetBstMidEndSepPunct{\mcitedefaultmidpunct}
{\mcitedefaultendpunct}{\mcitedefaultseppunct}\relax
\EndOfBibitem
\bibitem[Verstraete \emph{et~al.}(2023)Verstraete, Nishino, Schollw{\"o}ck, Ba{\~n}uls, Chan, and Stoudenmire]{verstraete2023density}
F.~Verstraete, T.~Nishino, U.~Schollw{\"o}ck, M.~C. Ba{\~n}uls, G.~K. Chan and M.~E. Stoudenmire, \emph{Nature Reviews Physics}, 2023, \textbf{5}, 273--276\relax
\mciteBstWouldAddEndPuncttrue
\mciteSetBstMidEndSepPunct{\mcitedefaultmidpunct}
{\mcitedefaultendpunct}{\mcitedefaultseppunct}\relax
\EndOfBibitem
\bibitem[Stoudenmire and White(2012)]{stoudenmire2012studying}
E.~M. Stoudenmire and S.~R. White, \emph{Annu. Rev. Condens. Matter Phys.}, 2012, \textbf{3}, 111--128\relax
\mciteBstWouldAddEndPuncttrue
\mciteSetBstMidEndSepPunct{\mcitedefaultmidpunct}
{\mcitedefaultendpunct}{\mcitedefaultseppunct}\relax
\EndOfBibitem
\bibitem[White and Martin(1999)]{white1999ab}
S.~R. White and R.~L. Martin, \emph{The Journal of chemical physics}, 1999, \textbf{110}, 4127--4130\relax
\mciteBstWouldAddEndPuncttrue
\mciteSetBstMidEndSepPunct{\mcitedefaultmidpunct}
{\mcitedefaultendpunct}{\mcitedefaultseppunct}\relax
\EndOfBibitem
\bibitem[Mitrushenkov \emph{et~al.}(2001)Mitrushenkov, Fano, Ortolani, Linguerri, and Palmieri]{mitrushenkov2001quantum}
A.~O. Mitrushenkov, G.~Fano, F.~Ortolani, R.~Linguerri and P.~Palmieri, \emph{The Journal of Chemical Physics}, 2001, \textbf{115}, 6815--6821\relax
\mciteBstWouldAddEndPuncttrue
\mciteSetBstMidEndSepPunct{\mcitedefaultmidpunct}
{\mcitedefaultendpunct}{\mcitedefaultseppunct}\relax
\EndOfBibitem
\bibitem[Chan and Head-Gordon(2002)]{chan2002highly}
G.~K.-L. Chan and M.~Head-Gordon, \emph{The Journal of chemical physics}, 2002, \textbf{116}, 4462--4476\relax
\mciteBstWouldAddEndPuncttrue
\mciteSetBstMidEndSepPunct{\mcitedefaultmidpunct}
{\mcitedefaultendpunct}{\mcitedefaultseppunct}\relax
\EndOfBibitem
\bibitem[Legeza \emph{et~al.}(2003)Legeza, R{\"o}der, and Hess]{legeza2003controlling}
{\"O}.~Legeza, J.~R{\"o}der and B.~Hess, \emph{Physical Review B}, 2003, \textbf{67}, 125114\relax
\mciteBstWouldAddEndPuncttrue
\mciteSetBstMidEndSepPunct{\mcitedefaultmidpunct}
{\mcitedefaultendpunct}{\mcitedefaultseppunct}\relax
\EndOfBibitem
\bibitem[Larsson \emph{et~al.}(2022)Larsson, Zhai, Umrigar, and Chan]{larsson2022chromium}
H.~R. Larsson, H.~Zhai, C.~J. Umrigar and G.~K.-L. Chan, \emph{Journal of the American Chemical Society}, 2022, \textbf{144}, 15932--15937\relax
\mciteBstWouldAddEndPuncttrue
\mciteSetBstMidEndSepPunct{\mcitedefaultmidpunct}
{\mcitedefaultendpunct}{\mcitedefaultseppunct}\relax
\EndOfBibitem
\bibitem[Baiardi and Reiher(2020)]{baiardi2020density}
A.~Baiardi and M.~Reiher, \emph{The Journal of Chemical Physics}, 2020, \textbf{152}, 040903\relax
\mciteBstWouldAddEndPuncttrue
\mciteSetBstMidEndSepPunct{\mcitedefaultmidpunct}
{\mcitedefaultendpunct}{\mcitedefaultseppunct}\relax
\EndOfBibitem
\bibitem[Or{\'u}s(2019)]{orus2019tensor}
R.~Or{\'u}s, \emph{Nature Reviews Physics}, 2019, \textbf{1}, 538--550\relax
\mciteBstWouldAddEndPuncttrue
\mciteSetBstMidEndSepPunct{\mcitedefaultmidpunct}
{\mcitedefaultendpunct}{\mcitedefaultseppunct}\relax
\EndOfBibitem
\bibitem[O'Rourke \emph{et~al.}(2018)O'Rourke, Li, and Chan]{o2018efficient}
M.~J. O'Rourke, Z.~Li and G.~K.-L. Chan, \emph{Physical Review B}, 2018, \textbf{98}, 205127\relax
\mciteBstWouldAddEndPuncttrue
\mciteSetBstMidEndSepPunct{\mcitedefaultmidpunct}
{\mcitedefaultendpunct}{\mcitedefaultseppunct}\relax
\EndOfBibitem
\bibitem[Becca and Sorella(2017)]{becca2017quantum}
F.~Becca and S.~Sorella, \emph{Quantum Monte Carlo approaches for correlated systems}, Cambridge University Press, 2017\relax
\mciteBstWouldAddEndPuncttrue
\mciteSetBstMidEndSepPunct{\mcitedefaultmidpunct}
{\mcitedefaultendpunct}{\mcitedefaultseppunct}\relax
\EndOfBibitem
\bibitem[Luo and Clark(2019)]{luo2019backflow}
D.~Luo and B.~K. Clark, \emph{Physical review letters}, 2019, \textbf{122}, 226401\relax
\mciteBstWouldAddEndPuncttrue
\mciteSetBstMidEndSepPunct{\mcitedefaultmidpunct}
{\mcitedefaultendpunct}{\mcitedefaultseppunct}\relax
\EndOfBibitem
\bibitem[Hermann \emph{et~al.}(2020)Hermann, Sch{\"a}tzle, and No{\'e}]{hermann2020deep}
J.~Hermann, Z.~Sch{\"a}tzle and F.~No{\'e}, \emph{Nature Chemistry}, 2020, \textbf{12}, 891--897\relax
\mciteBstWouldAddEndPuncttrue
\mciteSetBstMidEndSepPunct{\mcitedefaultmidpunct}
{\mcitedefaultendpunct}{\mcitedefaultseppunct}\relax
\EndOfBibitem
\bibitem[Pfau \emph{et~al.}(2020)Pfau, Spencer, Matthews, and Foulkes]{pfau2020ab}
D.~Pfau, J.~S. Spencer, A.~G. Matthews and W.~M.~C. Foulkes, \emph{Physical review research}, 2020, \textbf{2}, 033429\relax
\mciteBstWouldAddEndPuncttrue
\mciteSetBstMidEndSepPunct{\mcitedefaultmidpunct}
{\mcitedefaultendpunct}{\mcitedefaultseppunct}\relax
\EndOfBibitem
\bibitem[Hermann \emph{et~al.}(2023)Hermann, Spencer, Choo, Mezzacapo, Foulkes, Pfau, Carleo, and No{\'e}]{hermann2023ab}
J.~Hermann, J.~Spencer, K.~Choo, A.~Mezzacapo, W.~M.~C. Foulkes, D.~Pfau, G.~Carleo and F.~No{\'e}, \emph{Nature Reviews Chemistry}, 2023, \textbf{7}, 692--709\relax
\mciteBstWouldAddEndPuncttrue
\mciteSetBstMidEndSepPunct{\mcitedefaultmidpunct}
{\mcitedefaultendpunct}{\mcitedefaultseppunct}\relax
\EndOfBibitem
\bibitem[Kent \emph{et~al.}(2020)Kent, Annaberdiyev, Benali, Bennett, Landinez~Borda, Doak, Hao, Jordan, Krogel, Kyl{\"a}np{\"a}{\"a},\emph{et~al.}]{kent2020qmcpack}
P.~R. Kent, A.~Annaberdiyev, A.~Benali, M.~C. Bennett, E.~J. Landinez~Borda, P.~Doak, H.~Hao, K.~D. Jordan, J.~T. Krogel, I.~Kyl{\"a}np{\"a}{\"a} \emph{et~al.}, \emph{The Journal of chemical physics}, 2020, \textbf{152}, 174105\relax
\mciteBstWouldAddEndPuncttrue
\mciteSetBstMidEndSepPunct{\mcitedefaultmidpunct}
{\mcitedefaultendpunct}{\mcitedefaultseppunct}\relax
\EndOfBibitem
\bibitem[Motta and Zhang(2018)]{motta2018ab}
M.~Motta and S.~Zhang, \emph{Wiley Interdisciplinary Reviews: Computational Molecular Science}, 2018, \textbf{8}, e1364\relax
\mciteBstWouldAddEndPuncttrue
\mciteSetBstMidEndSepPunct{\mcitedefaultmidpunct}
{\mcitedefaultendpunct}{\mcitedefaultseppunct}\relax
\EndOfBibitem
\bibitem[Haghshenas \emph{et~al.}(2021)Haghshenas, Cui, and Chan]{haghshenas2021numerical}
R.~Haghshenas, Z.-H. Cui and G.~K.-L. Chan, \emph{Physical Review Research}, 2021, \textbf{3}, 023057\relax
\mciteBstWouldAddEndPuncttrue
\mciteSetBstMidEndSepPunct{\mcitedefaultmidpunct}
{\mcitedefaultendpunct}{\mcitedefaultseppunct}\relax
\EndOfBibitem
\bibitem[Pople(1999)]{pople1999nobel}
J.~A. Pople, \emph{Reviews of Modern Physics}, 1999, \textbf{71}, 1267\relax
\mciteBstWouldAddEndPuncttrue
\mciteSetBstMidEndSepPunct{\mcitedefaultmidpunct}
{\mcitedefaultendpunct}{\mcitedefaultseppunct}\relax
\EndOfBibitem
\bibitem[Sandala and Noodleman(2011)]{sandala2011modeling}
G.~M. Sandala and L.~Noodleman, \emph{Nitrogen fixation: methods and protocols}, 2011,  293--312\relax
\mciteBstWouldAddEndPuncttrue
\mciteSetBstMidEndSepPunct{\mcitedefaultmidpunct}
{\mcitedefaultendpunct}{\mcitedefaultseppunct}\relax
\EndOfBibitem
\bibitem[Lee \emph{et~al.}(2023)Lee, Lee, Zhai, Tong, Dalzell, Kumar, Helms, Gray, Cui, Liu, Kastoryano, Babbush, Preskill, Reichman, Campbell, Valeev, Lin, and Chan]{expadvantage}
S.~Lee, J.~Lee, H.~Zhai, Y.~Tong, A.~M. Dalzell, A.~Kumar, P.~Helms, J.~Gray, Z.-H. Cui, W.~Liu, M.~Kastoryano, R.~Babbush, J.~Preskill, D.~R. Reichman, E.~T. Campbell, E.~F. Valeev, L.~Lin and G.~K.-L. Chan, \emph{Nature Communications}, 2023, \textbf{14}, 1952\relax
\mciteBstWouldAddEndPuncttrue
\mciteSetBstMidEndSepPunct{\mcitedefaultmidpunct}
{\mcitedefaultendpunct}{\mcitedefaultseppunct}\relax
\EndOfBibitem
\bibitem[Chan and Head-Gordon(2003)]{chan2003exact}
G.~K.-L. Chan and M.~Head-Gordon, \emph{The Journal of Chemical Physics}, 2003, \textbf{118}, 8551--8554\relax
\mciteBstWouldAddEndPuncttrue
\mciteSetBstMidEndSepPunct{\mcitedefaultmidpunct}
{\mcitedefaultendpunct}{\mcitedefaultseppunct}\relax
\EndOfBibitem
\bibitem[Liakos and Neese(2015)]{liakos2015possible}
D.~G. Liakos and F.~Neese, \emph{Journal of chemical theory and computation}, 2015, \textbf{11}, 4054--4063\relax
\mciteBstWouldAddEndPuncttrue
\mciteSetBstMidEndSepPunct{\mcitedefaultmidpunct}
{\mcitedefaultendpunct}{\mcitedefaultseppunct}\relax
\EndOfBibitem
\bibitem[Liu \emph{et~al.}(2024)Liu, Du, Peng, Gray, and Chan]{liu2024tensor}
W.-Y. Liu, S.-J. Du, R.~Peng, J.~Gray and G.~K.-L. Chan, \emph{arXiv:2405.03797}, 2024\relax
\mciteBstWouldAddEndPuncttrue
\mciteSetBstMidEndSepPunct{\mcitedefaultmidpunct}
{\mcitedefaultendpunct}{\mcitedefaultseppunct}\relax
\EndOfBibitem
\bibitem[Liu \emph{et~al.}(2024)Liu, Zhai, Peng, Gu, and Chan]{liu2024hubbard}
W.-Y. Liu, H.~Zhai, R.~Peng, Z.-C. Gu and G.~K.-L. Chan, \emph{arXiv}, 2024\relax
\mciteBstWouldAddEndPuncttrue
\mciteSetBstMidEndSepPunct{\mcitedefaultmidpunct}
{\mcitedefaultendpunct}{\mcitedefaultseppunct}\relax
\EndOfBibitem
\bibitem[Liu \emph{et~al.}(2021)Liu, Huang, Gong, and Gu]{liu2021accurate}
W.-Y. Liu, Y.-Z. Huang, S.-S. Gong and Z.-C. Gu, \emph{Physical Review B}, 2021, \textbf{103}, 235155\relax
\mciteBstWouldAddEndPuncttrue
\mciteSetBstMidEndSepPunct{\mcitedefaultmidpunct}
{\mcitedefaultendpunct}{\mcitedefaultseppunct}\relax
\EndOfBibitem
\bibitem[Chen \emph{et~al.}(2024)Chen, Huang, Preskill, and Zhou]{chen2024local}
C.-F. Chen, H.-Y. Huang, J.~Preskill and L.~Zhou, Proceedings of the 56th Annual ACM Symposium on Theory of Computing, 2024, pp. 1323--1330\relax
\mciteBstWouldAddEndPuncttrue
\mciteSetBstMidEndSepPunct{\mcitedefaultmidpunct}
{\mcitedefaultendpunct}{\mcitedefaultseppunct}\relax
\EndOfBibitem
\bibitem[Peruzzo \emph{et~al.}(2014)Peruzzo, McClean, Shadbolt, Yung, Zhou, Love, Aspuru-Guzik, and O’Brien]{peruzzo2014variational}
A.~Peruzzo, J.~McClean, P.~Shadbolt, M.-H. Yung, X.-Q. Zhou, P.~J. Love, A.~Aspuru-Guzik and J.~L. O’Brien, \emph{Nature communications}, 2014, \textbf{5}, 4213\relax
\mciteBstWouldAddEndPuncttrue
\mciteSetBstMidEndSepPunct{\mcitedefaultmidpunct}
{\mcitedefaultendpunct}{\mcitedefaultseppunct}\relax
\EndOfBibitem
\bibitem[Motta \emph{et~al.}(2020)Motta, Sun, Tan, O’Rourke, Ye, Minnich, Brandao, and Chan]{motta2020determining}
M.~Motta, C.~Sun, A.~T. Tan, M.~J. O’Rourke, E.~Ye, A.~J. Minnich, F.~G. Brandao and G.~K.-L. Chan, \emph{Nature Physics}, 2020, \textbf{16}, 205--210\relax
\mciteBstWouldAddEndPuncttrue
\mciteSetBstMidEndSepPunct{\mcitedefaultmidpunct}
{\mcitedefaultendpunct}{\mcitedefaultseppunct}\relax
\EndOfBibitem
\bibitem[Bauer \emph{et~al.}(2020)Bauer, Bravyi, Motta, and Chan]{bauer2020quantum}
B.~Bauer, S.~Bravyi, M.~Motta and G.~K.-L. Chan, \emph{Chemical Reviews}, 2020, \textbf{120}, 12685--12717\relax
\mciteBstWouldAddEndPuncttrue
\mciteSetBstMidEndSepPunct{\mcitedefaultmidpunct}
{\mcitedefaultendpunct}{\mcitedefaultseppunct}\relax
\EndOfBibitem
\bibitem[Peng \emph{et~al.}(2023)Peng, Zhang, and Chan]{peng2023fermionic}
L.~Peng, X.~Zhang and G.~K.-L. Chan, \emph{Journal of Chemical Theory and Computation}, 2023, \textbf{19}, 9151--9160\relax
\mciteBstWouldAddEndPuncttrue
\mciteSetBstMidEndSepPunct{\mcitedefaultmidpunct}
{\mcitedefaultendpunct}{\mcitedefaultseppunct}\relax
\EndOfBibitem
\bibitem[Larocca \emph{et~al.}(2024)Larocca, Thanasilp, Wang, Sharma, Biamonte, Coles, Cincio, McClean, Holmes, and Cerezo]{larocca2024review}
M.~Larocca, S.~Thanasilp, S.~Wang, K.~Sharma, J.~Biamonte, P.~J. Coles, L.~Cincio, J.~R. McClean, Z.~Holmes and M.~Cerezo, \emph{arXiv preprint arXiv:2405.00781}, 2024\relax
\mciteBstWouldAddEndPuncttrue
\mciteSetBstMidEndSepPunct{\mcitedefaultmidpunct}
{\mcitedefaultendpunct}{\mcitedefaultseppunct}\relax
\EndOfBibitem
\bibitem[Cerezo \emph{et~al.}(2023)Cerezo, Larocca, Garc{\'\i}a-Mart{\'\i}n, Diaz, Braccia, Fontana, Rudolph, Bermejo, Ijaz, Thanasilp,\emph{et~al.}]{cerezo2023does}
M.~Cerezo, M.~Larocca, D.~Garc{\'\i}a-Mart{\'\i}n, N.~L. Diaz, P.~Braccia, E.~Fontana, M.~S. Rudolph, P.~Bermejo, A.~Ijaz, S.~Thanasilp \emph{et~al.}, \emph{arXiv preprint arXiv:2312.09121}, 2023\relax
\mciteBstWouldAddEndPuncttrue
\mciteSetBstMidEndSepPunct{\mcitedefaultmidpunct}
{\mcitedefaultendpunct}{\mcitedefaultseppunct}\relax
\EndOfBibitem
\bibitem[Haghshenas \emph{et~al.}(2022)Haghshenas, Gray, Potter, and Chan]{variationaltns}
R.~Haghshenas, J.~Gray, A.~C. Potter and G.~K.-L. Chan, \emph{Phys. Rev. X}, 2022, \textbf{12}, 011047\relax
\mciteBstWouldAddEndPuncttrue
\mciteSetBstMidEndSepPunct{\mcitedefaultmidpunct}
{\mcitedefaultendpunct}{\mcitedefaultseppunct}\relax
\EndOfBibitem
\bibitem[Nielsen and Chuang(2010)]{nielsen2010quantum}
M.~A. Nielsen and I.~L. Chuang, \emph{Quantum computation and quantum information}, Cambridge university press, 2010\relax
\mciteBstWouldAddEndPuncttrue
\mciteSetBstMidEndSepPunct{\mcitedefaultmidpunct}
{\mcitedefaultendpunct}{\mcitedefaultseppunct}\relax
\EndOfBibitem
\bibitem[Lin and Tong(2020)]{lin2020near}
L.~Lin and Y.~Tong, \emph{Quantum}, 2020, \textbf{4}, 372\relax
\mciteBstWouldAddEndPuncttrue
\mciteSetBstMidEndSepPunct{\mcitedefaultmidpunct}
{\mcitedefaultendpunct}{\mcitedefaultseppunct}\relax
\EndOfBibitem
\end{mcitethebibliography}
\bibliographystyle{rsc}
}

\end{document}